\documentclass[twocolumn,showpacs,preprintnumbers,amsmath,amssymb,showkeys]{
revtex4}

\usepackage{epsfig}
\usepackage[utf8]{inputenc}
\usepackage[T1]{fontenc}

\usepackage{graphicx}
\usepackage{dcolumn}
\usepackage{bm}
\usepackage{color}
\usepackage[active]{srcltx}

\newcommand{\cb}{\overline c}

\begin{document}

\title{Three-point correlation functions in Yang-Mills theory}

\author{Marcela Pel\'aez$^{ab}$}

\author{Matthieu Tissier$^a$}

\author{Nicol\'as Wschebor$^b$}

\affiliation{\vspace{.2cm}
$^a$LPTMC, Laboratoire de Physique Th\'eorique de la Mati\`ere 
Condens\'ee, CNRS UMR 7600, Universit\'e Pierre et Marie Curie, \\ boite 121, 4 
pl. Jussieu, 75252 Paris Cedex 05, France.\\
$^b$Instituto de F\'{\i}sica, Facultad de Ingenier\'{\i}a, Universidad de la Rep\'ublica,\\
J.~H. y Reissig 565, 11000 Montevideo, Uruguay.}

\date{\today}

\begin{abstract}
  We investigate the three-point correlation functions of Yang-Mills
  theory in the Landau gauge, with a particular emphasis on the
  infrared regime. The effect of the Gribov
  copies is accounted for by adding a mass term for the gluons in the
  Faddeev-Popov action in the Landau gauge. We perform a one-loop
  calculation for the ghost-antighost-gluon and three-gluon
  correlation functions. These analytic results are compared with the
  available lattice data and give a very satisfying agreement.
 \end{abstract}

\pacs{Valid PACS appear here}
\maketitle

\section{Introduction}
\label{sec_intro}

The physical observables of gauge theories are associated with
gauge-invariant quantities, which are therefore of utmost
interest. However in most of the analytical or semi-analytical
approaches, the determination of the expectation values of these
gauge-invariant quantities rely on understanding also the
gauge-dependent sector of the theory.  This is one of the reasons
why so much effort has been devoted to understand the properties of
correlation functions in Yang Mills theories in the past.

Several techniques are used to study these quantities. From the
analytic side, standard perturbation theory (that make use of the
Faddeev-Popov construction) is the most efficient tool to access the
ultraviolet regime of the theory, but fails at momenta of the order of
1~GeV. Indeed, the effective coupling (which is the expansion
parameter of perturbation theory) is large in this regime. The
standard perturbation theory even predicts that the coupling constant
diverges at the so-called infrared (IR) Landau pole.  For the
low-energy regime, the preferred analytical techniques are
nonperturbative renormalization-group and Schwinger-Dyson equation
methods \cite{vonSmekal97,Alkofer00,Zwanziger01,Fischer03,Bloch03,Pawlowski:2003hq,Fischer:2004uk,Aguilar04,Boucaud06,Aguilar07,Aguilar08,Boucaud08,Fischer08,RodriguezQuintero10,Huber:2012kd}.
These rely on a set of exact equations that are truncated, by
making some ansatz on a sector of the theory.

Another technique that has been used is lattice simulations that have played a central
role in our understanding of the correlation functions, in particular
in the Landau gauge which is rather easy to implement in
simulations. The extensive numerical work that has been performed in
the past decades
\cite{Cucchieri_08b,Cucchieri_08c,Cucchieri09,Bogolubsky09,Dudal10},
in conjunction with various semi-analytical techniques including various SD studies, 
  allowed to settle the controversy between two possible solutions of
  Schwinger-Dyson equations.  The
  so-called scaling solution corresponds to a gluon propagator that
  tend to zero at low momentum and a ghost dressing function (the
  propagator multiplied by the momentum squared) that diverges in this
  limit
  \cite{vonSmekal97,Alkofer00,Zwanziger01,Fischer03,Bloch03,Fischer08}. The so-called massive or decoupling
  solution gives a finite gluon propagator and a regular ghost
  dressing function at low momentum
  \cite{Bloch03,Aguilar04,Boucaud06,Aguilar07,Aguilar08,Boucaud08,RodriguezQuintero10,Huber:2012kd}. Lattice
  simulations clearly favored the second option in dimensions higher
  than two and the first one in the two-dimensional case
  \cite{Cucchieri_08c,Maas_07}.

In the past years, two of us have developed a new approach to access
the infrared behavior of the correlation functions. It relies on the
fact that, as is well known since the work of Gribov \cite{Gribov77}, the
Faddeev-Popov construction which is at the heart of most of the
analytical approaches, is not fully justified \footnote{Another approach to
study the impact of Gribov copies in the infrared sector of Yang-Mills
theory has been followed for many years, mainly by Zwanziger \cite{Zwanziger89,Zwanziger92,Zwanziger01,Dudal08}. Different
approximations performed in that scheme gave propagators compatible
with both massive and scaling scenarios.}. This is due to the fact
that this procedure does not completely fix the gauge. This so-called
Gribov ambiguity is however known to be unimportant in the high
momentum regime, and is only susceptible of modifying the infrared
properties. The idea pushed forward
in \cite{Tissier:2010ts,Tissier:2011ey} consists in modeling the
influence of the Gribov copies by adding a mass term for the gluons to
the usual Faddeev-Popov action (this leads to the Curci-Ferrari model
in the Landau gauge \cite{Curci76}). This idea was made more precise
in \cite{Serreau:2012cg} where a new Landau gauge-fixing was proposed,
which takes into account the Gribov ambiguity from first principles
and which leads, as far as perturbation theory is concerned, to the
same results as those obtained with the massive extension considered
in \cite{Tissier:2010ts,Tissier:2011ey}.
A one-loop calculation for the
gluon propagator and the ghost dressing function was performed that compared very well
with lattice simulations in $d=4$, with a
maximum error of $\sim$ 10$\%$, both for $SU(2)$ and $SU(3)$. 

It may look surprising that the infrared (often called
nonperturbative) regime of the theory can be reproduced to that level
of precision with a modest one-loop calculation. Our interpretation of
this fact is that the mass regularizes the theory in the infrared. For
example, we found renormalization schemes were the Landau pole
disappears. Moreover we made in \cite{Tissier:2011ey} an estimate  showing that,
in the infrared regime, the loop corrections to the propagators are
suppressed by powers of the external momenta (in particular, the 2-loop
corrections are roughly an order of magnitude smaller than the one-loop
contributions), which indicates that the perturbation theory {\it in presence
of a mass term} may be under control.

These encouraging results naturally lead us to consider other correlation
  functions. In consequence, in this article, we generalize our
  previous work to 3-point correlation functions. These functions are
  extremely interesting for many reasons:
  \begin{itemize}
   \item They
  have been calculated in lattice simulations
  \cite{Parrinello:1994wd,Alles:1996ka,Boucaud:1998bq,Cucchieri:2004sq,Ilgenfritz:2006gp,Cucchieri:2006tf,Cucchieri08} and
  this offers concrete data to compare with. 
  \item Once the parameters of the model have been fixed for the 2-point
  functions, the calculation of 3-point functions becomes a {\it pure
    prediction} without any parameter to adjust. This then becomes a
  very challenging test of the scheme proposed in \cite{Tissier:2010ts,Tissier:2011ey}.
  \item These functions
  are much richer than 2-point functions. In particular, instead of
  depending on a single momentum, they depend on three independent
  momenta squared. Moreover, they include various tensorial structures
  that could make their study even richer (even if these structures
  have not yet been studied  by lattice simulations).
  \end{itemize}

  In the past few years, several works aimed at describing these
  3-point correlation functions in the infrared regime, both with
  lattice simulations
  \cite{Parrinello:1994wd,Alles:1996ka,Boucaud:1998bq,Cucchieri:2004sq,Ilgenfritz:2006gp,Cucchieri:2006tf,Cucchieri08}
  and with semi-analytical methods
  \cite{RodriguezQuintero:2011au,Dudal:2012zx,Huber:2012kd,Aguilar:2013xqa}.
  However, the complexity of the standard semi-analytical methods (as
  Schwinger-Dyson equations or Non-Perturbative Renormalization Group
  equations) have delayed their study. In particular very few results
  on the ghost-antighost-gluon vertex are available
  \cite{RodriguezQuintero:2011au,Dudal:2012zx,Huber:2012kd,Aguilar:2013xqa}
  and essentially only models for the 3-gluon vertex have been
  proposed (see, for example, \cite{Huber:2012kd}).  On the contrary,
  the scheme developed in \cite{Tissier:2010ts,Tissier:2011ey} and
  that we follow here relies on a standard and simple 1-loop
  calculation.  Our main aim in this article is to show that we obtain
  3-point vertex functions which reproduce very well the lattice data
  for a relatively small computational effort. Note that Gracey
  \cite{Gracey:2012wf} studied the power corrections to the
  perturbative ultraviolet behaviour both in the model considered here
  and in the refined Gribov-Zwanziger model. These corrections are
  different so that lattice simulations with good precision would
  allow to make a definite difference between both methods.
  
  The outline of the article is the following. In
  Section~\ref{sec_oneloop}, we describe in more details the model and
  present our one-loop calculation. We then describe in
  Section~\ref{sec_renorm} the renormalization schemes that we
  implemented and finally describe our results and compare them to the
  lattice data available for the 3-point correlation functions in
  Section~\ref{sec_res}. We give our conclusions in
  Section~\ref{sec_conc}.

\section{One-loop calculation}
\label{sec_oneloop}

Our starting point is the Curci-Ferrari action in the Landau gauge,
written in Euclidean space, that reads:
\begin{equation}
  \label{eq_action}
  \begin{split}
      S=\int d^dx&\left[\frac 14 F_{\mu\nu}^aF_{\mu\nu}^a+ih^a\partial_\mu
    A_\mu^a +\partial_\mu\cb^a(D_\mu c)^a\right.\\
&\qquad\left.+\frac 12 m_0^2 (A_\mu^a)^2\right]
  \end{split}
\end{equation}
where the covariant derivative applied to a field $X$ in the adjoint
representation reads $(D_\mu X)^a=\partial_ \mu X^a+g_0f^{abc}A_\mu^b
X^c$, $g_0$ is the bare coupling constant, $f^{abc}$ are the structure
constants of the gauge group and
$F_{\mu\nu}^a=\partial_\mu A_\nu^a -\partial_\nu A_\mu^a+ g_0
f^{abc}A_\mu^a A_\nu ^b$ is the field strength. Apart for the bare
mass $m_0$ of the gluons, the action is that of the Yang-Mills theory
with the Faddeev-Popov action in the Landau gauge. All our
analytical calculations will be done for a generic $SU(N)$ gauge group.

The Feynman rules are the standard ones, except for the free propagator of
the gluon which reads:
\begin{equation}
  \label{eq_propag_AA}
  \langle
  A_\mu^aA_\nu^b\rangle_0(p)=\delta^{ab} P^\perp_{\mu\nu}(p)\frac 1{p^2+m_0^2}
\end{equation}
where we introduced the transverse projector (and, for later use, the
longitudinal one):
\begin{align}
  P^\perp_{\mu\nu}(p)&=\delta_{\mu\nu}-\frac{p_\mu p_\nu}{p^2}\\
  P^\parallel_{\mu\nu}(p)&=\frac{p_\mu p_\nu}{p^2}
\end{align}

Instead of computing the correlation functions, we compute as
usual the vertex functions which are obtained by considering only the
one-particle irreducible (1PI) diagrams. We parametrize the two-point vertex functions in terms of three scalar
functions:
\begin{align}
&  
\Gamma^{(2)}_{A_\mu^aA_\nu^b}(p)=\delta^{ab}\big(\Gamma^\perp(p)P^\perp_{\mu\nu}
+\Gamma^\parallel(p)P^\parallel_{\mu\nu}\big)\\
& \Gamma^{(2)}_{c^a \cb^b}(p)=\delta^{ab} \frac{p^2}{J(p)} 
\end{align}
The full propagators for the gluon and ghost then read:
\begin{align}
\langle  A_\mu^aA_\nu^b\rangle(p)=\delta^{ab} \frac{P^\perp_{\mu\nu}(p)}{\Gamma^\perp(p)}  \\
\langle  c^a\cb^b\rangle(p)=\delta^{ab} \frac{J(p)}{p^2 }  
\end{align}
The function $J$ is the so-called dressing functions. The 1-loop
expressions for $\Gamma^\perp$ and $J$ were computed in
\cite{Tissier:2010ts,Tissier:2011ey}. The longitudinal part of the gluon two-point
function is not directly accessible in lattice simulations and was
therefore not considered in the past. It proved however interesting to
compute it because it appears in some Ward identities that we used to
check the consistency of our results. The 1-loop expressions for
these three functions are given in the supplemental material
\cite{supplemental}.

\subsection{Ghost-antighost-gluon vertex}

\begin{figure}[tbp]
  \centering
  \includegraphics[width=.4\linewidth]{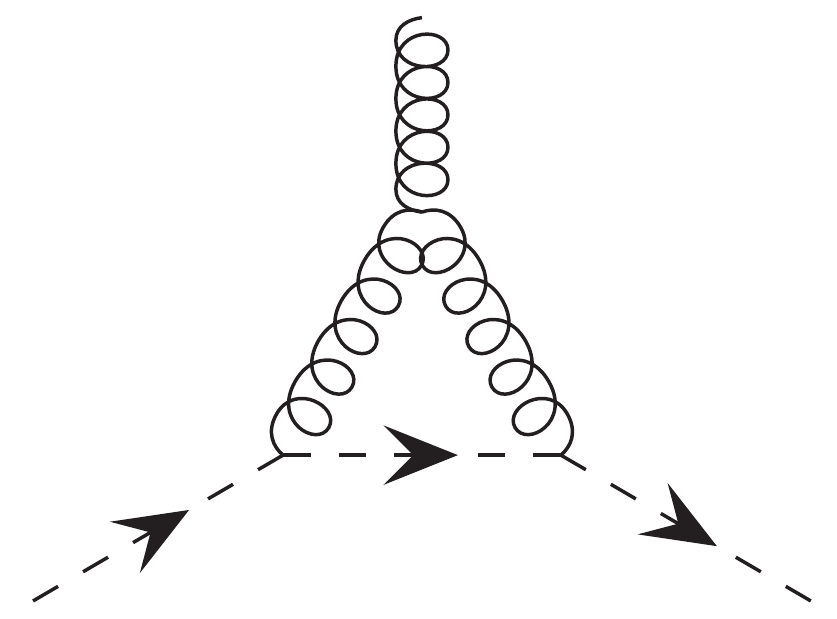}\hfill
  \includegraphics[width=.4\linewidth]{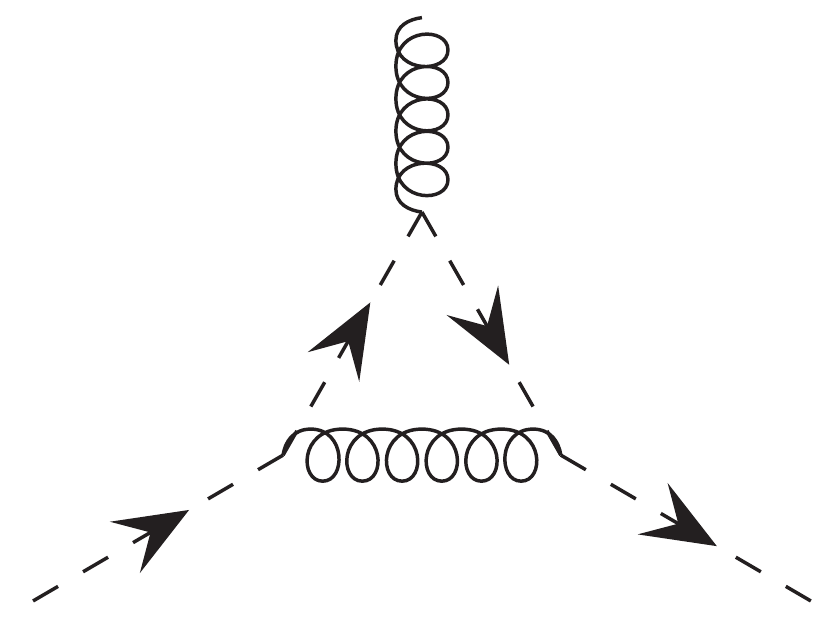}
  \caption{One-loop Feynman diagrams for the ghost-antighost-gluon
    vertex.}
  \label{fig_diagccbA}
\end{figure}

The one-loop calculation for the
ghost-antighost-gluon vertex function requires computing the two
Feynman diagrams showed in Fig.~\ref{fig_diagccbA}.
The tensorial structure is rather simple. At one loop, the vertex is
proportional to $f^{abc}$ where $a$, $b$ and $c$ are respectively the
color indices of the ghost, antighost and gluon external legs. 
We will consider here and below only this color structure. This is
an exact property for $SU(2)$, and for general $SU(N)$, it remains true at two loops and also in the large $N$
limit. However, this is only an approximation for $SU(3)$
(currently used in the literature \cite{BallChiu}).
The Lorentz index $\mu$ of the gluon external leg can be carried by one of
the external momenta. Because of the momentum conservation, there are
actually only two scalar components. Following Ball and Chiu
\cite{BallChiu}, it is however convenient to express the vertex in
terms of a rank two tensor $\Gamma_{\nu\mu}$ such that:
\begin{equation}
  \label{eq_vertex_cbcA_def}
  \Gamma^{(3)}_{c^a\cb^b A_\mu^c}(p,k,r)=-i g_0 f^{abc} k_{\nu}\Gamma_{\nu\mu}(p,k,r)
\end{equation}
with 
\begin{equation} 
\label{eq_vertex_KbcA_def}
  \begin{split}
  \Gamma_{\nu\mu}(p,k,r)&=\delta_{\mu\nu}a(r^2,k^2,p^2)\\
&-r_{\nu}p_{\mu}b(r^2,k^2,p^2)  +k_{\nu}r_{\mu}c(r^2,k^2,p^2)\\
&+r_{\nu}k_{\mu}d(r^2,k^2,p^2)+k_{\nu}k_{\mu}e(r^2,k^2,p^2)    
  \end{split}
\end{equation}
Note that lattice simulations access the vertex function only through the 
correlation function, {\it i.e.} the vertex function with external legs 
contracted with the full propagators. Since the gluon propagator is transverse, 
the function $c$ is not accessible to lattice simulations.
In fact, the lattice simulations on the 
ghost-antighost-gluon correlation function that have actually been performed \cite{Cucchieri08} are presented in terms of a scalar function 
which is obtained by contracting the external gluon leg with the transverse 
propagator and with the bare ghost-antighost-gluon vertex, normalized by the 
same expression at tree level:
\begin{equation}
\label{GccbA}
 G^{c\cb A}(p,k,r)=\frac{k_\nu P^\perp_{\mu\nu}(r)k_\rho 
\Gamma_{\rho\mu}(p,k,r)} {k_\nu P^\perp_{\mu\nu}(r)k_\mu }
\end{equation}
A simple 
calculation shows that this correlation function depends 
on a unique linear combination of the scalar functions defined in (\ref{eq_vertex_KbcA_def}):
\begin{equation}
a+k\cdot r (b+d)+k^2 e
\end{equation}
with the functions $a,b,d$ and $e$ evaluated at $(r^2,k^2,p^2)$. Although this is the only available
numerical data, it is interesting to
compute the 5 scalar functions $a,b,c,d$ and $e$ because this gives an
internal check of the validity of the calculation, see below. Moreover, future lattice
studies may lead to a determination of the various tensorial components independently.

In our calculations, we decomposed each diagram on the tensorial
structure of Eq.~(\ref{eq_vertex_KbcA_def}), following the ideas of
\cite{Passarino78}. We can then deduce the contribution of a diagram to each of 
the scalar functions $a$ to $e$ in terms of
integrals. To do so, it is convenient to rewrite the product of a
massive propagator and of a massless propagator as:
\begin{equation}
  \label{eq_urko}
\frac 1{p^2(p^2+m^2)}=\frac 1 {m^2}\left(\frac{1}{p^2}-\frac{1}{p^2+m^2}\right) 
\end{equation}
We can thus express the scalar functions in terms of a few simple
integrals. By using Feynman parameters, we can perform the momentum
integral and obtain expressions with at most one integral over a
Feynman parameter that cannot be performed analytically for generic
momentum configurations. The expression are lengthy and not
particularly instructive. We give them in the supplemental material
\cite{supplemental}. The calculation simplifies in the case  of one
vanishing external momentum. In particular, when the ghost or antighost momentum
vanishes, the vertex function have no loop corrections \cite{Taylor71}. For vanishing gluon
momentum, the vertex is non-trivial and we find, in $d=4-\epsilon$:
\begin{equation}
  \label{eq_1loopexceptional}
  \begin{split}
  \Gamma_{\mu\nu}(p,-p,&0)=\delta_{\mu\nu}\bigg\{1+\frac{g_0^2 
N}{128 \pi ^2}\bigg[9/2+s\\&+5s^{-1}-(7 s^{-1}+5s^{-2}) \log (s+1)\\&-(s-1) s 
\log \left(s^{-1}+1\right)\bigg]\bigg\}    
  \end{split}
\end{equation}
were we introduced $s=p^2/m^2$. Note that the previous (bare) vertex is finite
due to the non-renormalization theorem to be discussed below. The equivalent expression in $d=3$ reads:
\begin{equation}
  \label{eq_1loopexceptional_3d}
  \begin{split}
  \Gamma_{\mu\nu}(p,-p,&0)=
\delta_{\mu\nu}\bigg\{1+\frac{g_0^2 
N}{384 \pi m s}\bigg[2(6s^2\\&-5s-21)-3\pi\sqrt 
s(2s^2-s+288)\\&+6s^{-1/2}(2s^3-s^2-68s+7)\arctan(\sqrt{s})\bigg]\bigg\}    
  \end{split}
\end{equation}

\subsection{Three gluon vertex}

\begin{figure}[bp]
  \centering
  \includegraphics[width=.3\linewidth]{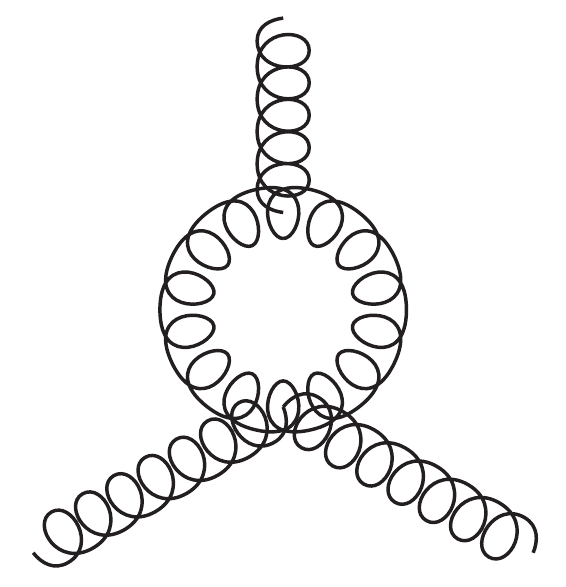}\hfill
  \includegraphics[width=.4\linewidth]{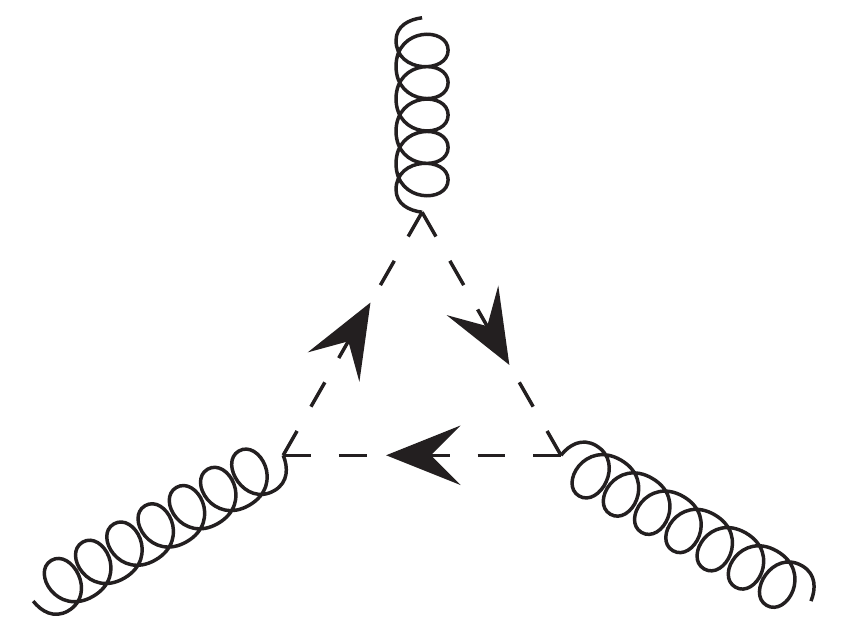}
  \includegraphics[width=.4\linewidth]{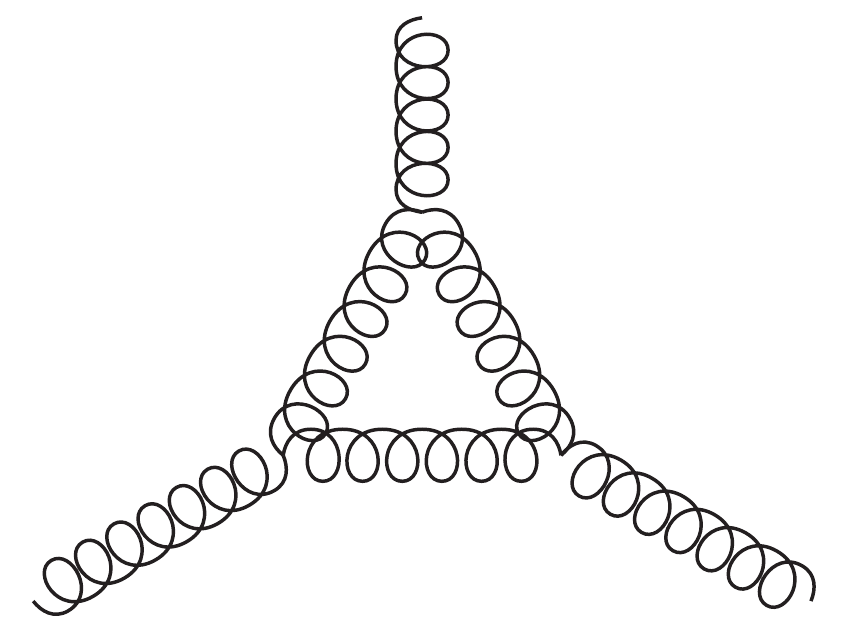}
  \caption{One-loop Feynman diagrams for the 3-gluon vertex.}
  \label{fig_diagAAA}
\end{figure}

The one-loop calculation for the three-gluon correlation
function requires computing the three Feynman diagrams showed in
Fig.~\ref{fig_diagAAA}.
Again, the color structure is rather simple at one loop, being
proportional to $f^{abc}$.  As for the ghost-antighost-gluon vertex, we will
ignore the possibility of more involved color structures (that are absent
at two-loops order, for $SU(2)$ gauge group and also in the large $N$ limit of $SU(N)$ gauge groups)
Accordingly, we define:
$$\Gamma_{A^a_\mu A^b_\nu A^c_\rho}^{(3)}(p,k,r)=-ig_0f^{abc}\Gamma_{\mu\nu\rho}(p,k,r).$$
The Lorentz structure is richer  than in the previous case since there are now three Lorentz
indices. We have used the decomposition of Ball and Chiu
\cite{BallChiu} to extract six scalar functions: 
\begin{widetext}
\begin{equation}
  \begin{split}
\Gamma_{\mu\nu\rho}(p,k,r)&=A(p^2,k^2,r^2)\delta_{\mu\nu}(p-k)_\rho+ 
B(p^2,k^2,r^2)\delta_{\mu\nu}(p+k)_\rho 
-C(p^2,k^2,r^2)(\delta_{\mu\nu}p.k-p_{\nu}k_{\mu})(p-k)_\rho\\
&+\frac{1}{3}S(p^2,k^2,r^2)(p_{\rho}k_{\mu}r_{\nu}+p_{\nu}k_{\rho}r_{\mu})+  
F(p^2,k^2,r^2)(\delta_{\mu\nu}p.k-p_{\nu}k_{\mu})(p_{\rho}k.r-k_{\rho}p.r)\\
&+H(p^2,k^2,r^2)\left[-\delta_{\mu\nu}(p_{\rho}k.r-k_{\rho}p.r)+\frac{1}{3}(p_{
\rho}k_{\mu}r_{\nu}-p_{\nu}k_{\rho}r_{\mu})\right]+\text{cyclic permutations}  
  \end{split}
\end{equation}
The scalar functions have the following symmetry properties: $A$, $C$
and $F$ are symmetric under permutation of the first two arguments;
$B$ is antisymmetric under permutation of the first two arguments; $H$
is completely symmetric and $S$ is completely antisymmetric. As for
the ghost-antighost-gluon vertex, only a subset of these functions are
measurable in lattice
simulations; when the external legs are contracted with the gluon
propagators (which is transverse), the functions $B$ and $S$
disappear. In lattice simulations \cite{Cucchieri08}, the quantity which 
  has been considered is a scalar function obtained by contracting
the external legs of the vertex with transverse propagators and the
tree-level momentum structure of the 3-gluon vertex, normalized to the
same expression at the bare level:
\begin{equation}
\label{GAAA}
G^{AAA}(p,k,r)=\frac{[(r-k)_\gamma\delta_{\alpha\beta}+\text{cyclic 
permutations}] P^\perp_{\alpha\mu}(p)P^\perp_{ \beta\nu } 
(k)P^\perp_ { \gamma\rho } 
(r)\Gamma_{\mu\nu\rho}(p,k,r)}{[(r-k)_\gamma\delta_{\alpha\beta} 
+\text{cyclic 
permutations}] 
P^\perp_{\alpha\mu}(p)P^\perp_{ \beta\nu } (k)P^\perp_ { \gamma\rho } 
(r)[(r-k)_\rho\delta_{\mu\nu}
+\text{cyclic 
permutations}]}
\end{equation} 

The scalar functions $A$, $B$, $C$, $S$, $F$ and $H$ are computed in
the same way as described above. Our expression for generic momenta
involve at most one integral over a Feynman parameter and cannot be
expressed in terms of elementary functions. They are given
in the supplemental material \cite{supplemental}. When one momentum
vanishes, the integral over the Feynman parameter can be performed 
analytically which simplifies considerably the results. For $d=4-\epsilon$:
\begin{equation}
\label{eq_3gluonexceptional}
\begin{split}
  \Gamma_{\mu\nu\rho}(p,0,-p)=& \bigg\{ 1 -\frac{N g_0^2}{768\pi^2 }\bigg[-\frac
  {136}{\epsilon}(1-\epsilon \log \bar m)+\frac 13(36 s^{-2}-594
  s^{-1}+319+6s)+(3s^2-2)\log  s\\
  &-4 s^{-3}(1+s)^3(s^2-9s+3)\log(1+s)\\
  &+\frac{(4+s)^{3/2}}{s^{3/2}}\left(24-30s+s^2\right) \log\left(\frac{\sqrt{4+s}+\sqrt
      s}{\sqrt{4+s}-\sqrt s}\right)\bigg]\bigg\} (p_\mu\delta_{\nu\rho}+p_\rho\delta_{\mu\nu})\\
  &- \bigg\{ 2 +\frac{N g_0^2}{384\pi^2 }\bigg[-\frac
  {136}{\epsilon}(1-\epsilon \log \bar m)+\frac 13(18 s^{-2}-321
  s^{-1}-97+24s)+(s-1)(s^2-2s-2)\log  s\\
  &-2s^{-3}(1+s)^2(s-1)(s^3-7s^2+7s-3)\log(1+s)\\
  &+\frac{\sqrt{4+s}}{s^{3/2}}\left(48+16
    s+22 s^2-11 s^3+s^4 \right) \log\left(\frac{\sqrt{4+s}+\sqrt
      s}{\sqrt{4+s}-\sqrt s}\right)\bigg]\bigg\} p_\nu\delta_{\mu\rho} \\  
  & -\frac{N g_0^2}{384 \pi^2 m^2}\bigg[(-36 s^{-3}+278
  s^{-2}-74 s^{-1}-10)-s^2\log  s\\
  &+ s^{-3}(1+s)^2(36s^{-1}-44-4s-12s^2+2s^3)\log(1+s)\\
  &+\sqrt{s(4+s)}s^{-3}(-144s^{-1}+80+4s+10s^2-s^3) \log\left(\frac{\sqrt{4+s}+\sqrt
      s}{\sqrt{4+s}-\sqrt s}\right)\bigg] p_\mu p_\nu p_\rho
\end{split}
\end{equation}
where $\bar m^2=m^2e^\gamma/(4\pi)$ with $\gamma$ the Euler
constant. The divergent term $\propto 1/\epsilon$ disappears 
once the renormalized vertex functions are expressed in 
terms of the renormalized parameters, see Sect.~\ref{sec_renorm}.

In $d=3$, the same quantity reads:
\begin{equation}
\label{eq_3gluonexceptional_3d}
\begin{split}
  \Gamma_{\mu\nu\rho}(p,0,-p)=&\bigg\{1 -\frac{N g_0^2}{128\pi m\sqrt s}\bigg[
  \frac\pi 2 (2-3s^2)+\frac{2}{3s^{3/2}}(3s^3+23s^2+56s-15)\\&
  -s^{-1}(4+s)(16-18s+s^2)\arctan(\sqrt s/2)
  +2s^{-2}(s-5)(s+1)^2\arctan(\sqrt s)\bigg]\bigg\} (p_\mu\delta_{\nu\rho}+p_\rho\delta_{\mu\nu}) \\
  & - \bigg\{ 2 +\frac{N g_0^2}{128\pi m\sqrt s}\bigg[
  \pi (-s^3+3s^2-1)+\frac{2}{3s^{3/2}}(15s^3-51s^2+53s-15)\\&+
  2(-s^3+6s^2+2s-16-32 s^{-1})\arctan(\sqrt s/2)
  +2(2 s^3 - 9 s^2 - 5)(1-s^{-2})\arctan(\sqrt s)\bigg]\bigg\} p_\nu\delta_{\mu\rho}\\
& -\frac{N g_0^2}{128\pi m^3 \sqrt s}\bigg[
  \pi  (s^{-1}+s^2)+\frac{2}{3s^{5/2}}(-21s^3+5s^2-139s+75)\\& +2 s^{-2}(s^4-5s^3-16s^2-40s+96)\arctan(\sqrt s/2) \\&
  -2s^{-3}(s+1)(2s^4-7s^3-9s^2-15s+25)\arctan(\sqrt s)\bigg] p_\mu p_\nu p_\rho
\end{split}
\end{equation}
\end{widetext}

\subsection{Checks}
We present in this section the different checks that can be performed
on our one-loop expressions.  First, as explained in the Appendix, the
vertex functions fulfill the following Slavnov-Taylor identity:
\begin{equation}
  \label{eq_BRST_constraint}
  \begin{split}
&[\Gamma^\perp(p)P^\perp_{\mu\rho}(p)+\Gamma^\parallel(p)P^\parallel_{\mu\rho}
(p)]\Gamma_{\mu\nu}(r,p,k)\\
&-[\Gamma^\perp(k)P^\perp_{\mu\nu}(k)+\Gamma^\parallel(k)P^\parallel_{\mu\nu}(k)
]\Gamma_{\mu\rho}(r,k,p)\\
&\qquad=r_\mu J^{-1}(r)\Gamma_{\rho\nu\mu}(p,k,r)
  \end{split}
\end{equation}
We have verified analytically that our one-loop expressions satisfy
this relation. This gives a nontrivial check for most of the scalar
functions, except however $F$ and $H$ which are associated with
transverse momentum structures and that do not appear in
Eq.~(\ref{eq_BRST_constraint}).

Second, we have compared our expressions in the limit of vanishing
mass with those of \cite{Davydychev}. Because we used the relation
(\ref{eq_urko}), there appear in our expressions terms in $1/m$ that
would naively diverge in the limit of small mass. We can check
explicitly that the limit is actually regular but the analytic
comparison is cumbersome. We have made instead a numerical comparison
of our expressions and those of \cite{Davydychev} for 50 momentum
configurations, taking the mass of the gluon much smaller than the
momenta. We have mainly considered the scalar functions appearing in
the ghost-antighost-gluon vertex (\ref{eq_vertex_KbcA_def}) and the
functions $F$ and $H$ that, as discussed above, are not constrained by
Eq.~(\ref{eq_BRST_constraint}). In all cases, our expressions in the
massless limit agree with \cite{Davydychev} at the numerical precision
level.

Finally we have considered the following equality, which is derived in the 
Appendix.
\begin{equation}
  \label{eq_non_renorm}
  \begin{split}
  \widetilde\Gamma_\mu(p,k,r)+ 
\widetilde\Gamma_\mu(k,p,r)-&\frac{r_\mu}{r^2}\bigg[ 
\frac{p_\nu}{p^2}\widetilde\Gamma_\nu(k,r,p)\\&+\frac{k_\nu}{k^2}
\widetilde\Gamma_\nu(p,r,k)\bigg]=    0
  \end{split}
\end{equation}
where
\begin{equation}
  \label{eq_gammatilde}
  \widetilde \Gamma_\mu(p,k,r)=k_\nu \Gamma_{\nu\mu}(p,k,r) r^2 J^{-1}(r)
\end{equation}
To our knowledge this relation have not been derived before.
We have checked numerically for 50 momentum configurations that the previous identity is indeed satisfied, with no
constraint on the mass.

\subsection{Infrared behavior}

It is instructive to discuss the behavior of the different vertex
functions when all the external momenta are much smaller than the mass
scale. A straightforward analysis shows that, in this limit, the
leading contribution comes from the diagram with as much ghost
propagators as possible. Multiplying all momenta by a common
coefficient $\kappa$, we obtain the following behaviors, valid in
arbitrary dimension:
\begin{equation}
\label{compIR}
 \begin{split}
  \Gamma_{\mu\nu}(\{\kappa p_i\})-\delta_{\mu\nu}&\sim \kappa^{d-2},\\
  \Gamma_{\mu\nu\rho}(\{\kappa p_i\})&\sim \kappa^{d-4}.
 \end{split}
\end{equation} 
As a consequence of these behavior, $G^{AAA}$ diverges as
$\log\kappa$ in $d=4$ and diverges as $1/\kappa$ in $d=3$ when $\kappa
\to 0$.

\section{Renormalization and Renormalization Group}
\label{sec_renorm}
 
In this section, we describe the renormalization schemes that we implemented
and explain how the renormalization-group ideas are implemented.
As we explain below, some care must be taken when comparing the (bare) lattice data with the
renormalized analytical results.

\subsection{Renormalization and schemes}
As usual, the
divergences appearing in the one-loop expressions can be absorbed
into a redefinition of the coupling constant, mass and fields. In $d=3$ no ultraviolet divergences are
present but a (finite) renormalization is done anyway in order to be able to exploit renormalization-group
methods and improve perturbation theory. We
define the renormalized quantities as:
\begin{align}
 A_0^{a\,\mu}= \sqrt{Z_A} A^{a\,\mu},&\hspace{.5cm} 
 c_0^{a}= \sqrt{Z_c} c^{a},\hspace{.5cm}
 \bar c_0^{a}= \sqrt{Z_c} \bar c^{a},\hspace{.5cm} \nonumber\\
g_0&= Z_g g \hspace{.5cm} m_0^2= Z_{m^2} m^2
\end{align}
From now on, except when explicitly stated, all quantities are the renormalized ones.
The relations between bare (with subindices ``0'') and renormalized vertices are the following:
\begin{align}
\label{renormvertices}
 \Gamma^{(2)}_{A_\mu^aA_\nu^b}(p)&=Z_A \Gamma^{(2)}_{A_\mu^aA_\nu^b,0}(p) \nonumber\\
 \Gamma^{(2)}_{c^a \cb^b}(p)&=Z_c  \Gamma^{(2)}_{c^a \cb^b,0}(p) \nonumber\\
 \Gamma^{(3)}_{c^a \cb^b A_\mu^c}(p,r)&=Z_c \sqrt{Z_A} \Gamma^{(3)}_{c^a \cb^b A_\mu^c,0}(p,r)\nonumber\\
 \Gamma^{(3)}_{A_\mu^aA_\nu^bA_\rho^c}(p,r)&=Z_A^{3/2} \Gamma^{(3)}_{A_\mu^aA_\nu^bA_\rho^c,0}(p,r) 
 \end{align}

We have used two renormalization schemes to fix the renormalization
factors, that were already presented in \cite{Tissier:2011ey}. The
vanishing-momentum (VM) scheme is characterized by 
\begin{align}
\label{eq_def_vanishing}
& \Gamma^\perp(p=\mu)=m^2+\mu^2, \hspace{.4cm}J(p=\mu)=1\nonumber\\
&\Gamma^\perp(p=0)=m^2.
\end{align}
The infrared safe scheme (IS) relies on a non-renormalization theorem
for the mass \cite{Dudal02,Wschebor07,Tissier08} (this
non-renormalization theorem was conjectured is \cite{Gracey:2002yt}),
which is imposed here for the finite part of the renormalization
parameters. It is defined by:
\begin{align}
\label{rencond}
&\Gamma^\perp(p=\mu)=m^2+\mu^2, \hspace{.4cm} J(p=\mu)=1,\nonumber\\
&Z_{m^2} Z_A Z_c=1.
\end{align}
In both cases, we use the Taylor scheme to fix the renormalization
factor of the coupling constant. This leads to:
\begin{equation}
  \label{eq_taylor}
  Z_g\sqrt{Z_A} Z_c=1
\end{equation}
The explicit expressions for the different renormalization factors are given in 
\cite{Tissier:2011ey}. 

It is important to relate the objects observed on lattice simulations $G_0^{c\cb A}$ and $G_0^{AAA}$
given by Eqs.~(\ref{GccbA}) and (\ref{GAAA}) to the renormalized vertices. The quantities that are used
on the lattice are bare vertices without renormalization factors. When expressed in terms of the renormalized
vertices as in Eq.~(\ref{renormvertices}), we obtain:
\begin{align}
\label{renorm3point}
  G_0^{c\cb A}(p,k,r)&= G^{c\cb A}(p,k,r)\nonumber\\
  G_0^{AAA}(p,k,r)&=\frac{Z_c}{Z_A} G^{AAA}(p,k,r)
 \end{align}
where the renormalized expressions correspond to those written in Eqs.~(\ref{GccbA}) and (\ref{GAAA}) but
with the corresponding renormalized vertices instead of the bare ones. In order to arrive to this result
we exploited the Taylor's non-renormalization theorem (\ref{eq_taylor}). Of course, the lattice results are regularized
and accordingly the factor $Z_c/Z_A$ is finite but it is necessary to include it when comparing our renormalized results
to those coming from lattice simulations.

\subsection{Renormalization Group}
Once the correlation functions for the renormalized field are expressed in 
terms of the renormalized coupling constant and renormalized mass, we get 
finite expressions both in $d=3$ and in $d=4$. The direct comparison of these expressions with the lattice 
results was not completely satisfactory at energies of a few GeV. This is to 
be attributed to large loop corrections (in $d=4$ large logarithms $\propto \log(p/\mu)$)
and we therefore had 
to use a renormalization-group improvement of our one-loop expressions. To do 
so, we introduce the $\beta$ function and anomalous dimensions of the fields as:
\begin{align}
\beta_g(g,m^2)&=\mu\frac{dg}{d\mu}\Big|_{g_0, m^2_0},\\
\beta_{m^2}(g,m^2)&=\mu\frac{dm^2}{d\mu}\Big|_{g_0, m^2_0},\\
\gamma_A(g,m^2)&=\mu\frac{d\log Z_A}{d\mu}\Big|_{g_0, m^2_0},\\
\gamma_c(g,m^2)&=\mu\frac{d\log Z_c}{d\mu}\Big|_{g_0, m^2_0}.
\end{align}
We can then use the RG equation for the vertex function with $n_A$ gluon 
legs and $n_c$ ghost legs:
\begin{equation}
\begin{split}
\Big( \mu \partial_\mu -\frac 1 2 &(n_A \gamma_A+n_c \gamma_c)\\&+\beta_g 
\partial_{g}+
\beta_{m^2}\partial_{m^2}\Big)\Gamma^{(n_A,n_c)}=0,
\end{split}
\end{equation}
to relate these functions at different scales:
\begin{equation}
\label{eq_int_RG}
\begin{split}
\Gamma^{(n_A,n_c)}(&\{p_i\},\mu,g(\mu),m^2(\mu))=z_A(\mu)^{
n_A/2}z_c(\mu)^{
n_c/2}\\
&\times\Gamma^{(n_A,n_c)}(\{p_i\},\mu_0,g(\mu_0),m^2(\mu_0)).
\end{split} 
\end{equation} 
where $g(\mu)$ and $m^2(\mu)$ are obtained by integration of the
beta functions with initial conditions given at some scale $\mu_0$ and:  
\begin{equation}
\label{eq_def_z_phi}
\begin{split}
\log z_A(\mu)&=\int_{\mu_0}^\mu\frac
     {d\mu'}{\mu'}\gamma_A\left(g(\mu'),m^2(\mu')\right),\\ \log
     z_c(\mu)&=\int_{\mu_0}^\mu\frac
     {d\mu'}{\mu'}\gamma_c\left(g(\mu'),m^2(\mu')\right).
\end{split}
\end{equation}

There remains to choose the RG scale $\mu$ at which Eq.~(\ref{eq_int_RG}) is 
evaluated. For a correlation function with typical momentum $p$, in the 
UV regime $p\gg m$, it is important to take $\mu\simeq p$. However, in the IR 
regime, the theory is effectively massive and no large logarithm are present. 
It is therefore not necessary to integrate the flow down to RG 
scales smaller than $m$. We therefore used a running scale: \footnote{A 
conceptually more satisfying criterion would be to use the mass at the running 
scale $\mu$ instead of $\mu_0$. However, the variation of the mass is small in 
the range of energy of interest for us so that the difference between these two 
schemes is tiny.}
\begin{equation}
 \label{eq_rg_scale}
 \mu=\sqrt{p^2+\alpha\, m^2(\mu_0)}
\end{equation} 
where $\alpha$ is a parameter that, in principle, can vary between zero and values of order one.
In practice we used various values of
$\alpha$ between 0 and 3. We discuss for the two schemes and for $d=4$ and $d=3$ the dependence on $\alpha$ in the following section.

\section{Results}
\label{sec_res}

In this section, we present the results for the 2 and 3-point
functions both in $d=4$ and in $d=3$. As explained in
\cite{Tissier:2010ts,Tissier:2011ey} the perturbative scheme considered here
does not work in $d=2$. In that case, the loop corrections contains
infrared divergences that a IR Landau-pole remains present, even if a
mass for the gluons have been introduced. For this reason, we only
present results for higher dimensions.

\subsection{Fixing parameters}

Following \cite{Tissier:2010ts,Tissier:2011ey}, we consider here the values of the
mass and coupling constants at some renormalization
scale $\mu_0$ as fitting parameters. Since the lattice results
are much more precise for the propagators than for the
3-point correlation functions, we look for the set of parameters
that lead to the best fit of the 2-point correlation functions.
[Note that
the global normalization of the ghost and gluon two-point correlation functions
are not accessible so we have to
introduce a multiplicative factor in front of our analytic expression
of the propagators, which must be fixed by comparison with the lattice data.]
The values of the mass and coupling constant
thus determined are then used for computing the 3-point
correlation function. When comparing the latter with lattice results,
we are thus left with only one free parameter associated with the
global normalization of $G^{AAA}$.
For each lattice parameter beta, we fixed this multiplicative factor
by considering a particular momentum configuration
(with one vanishing momentum) and used the same value for the other
momentum configurations \footnote{In principle, this normalization of $G^{AAA}$ could be fixed by
normalizing the lattice results with the gluon and ghost propagators
in such a way that the continuum limit is well defined. We would then
have one less parameter to fix when comparing with lattice data. This would
give a more stringent test of theoretical results.}. In previous work
we compared both the $SU(2)$ and the $SU(3)$ cases with lattice results \cite{Tissier:2010ts,Tissier:2011ey}. In the present
article, given that the available lattice simulations for the infrared behaviour of
3-point functions are for the $SU(2)$ group, we only present numerical results for that case.

When comparing lattice data with analytical results,
it is important to have simultaneously a small relative and absolute error
since the propagators tend to zero in the ultraviolet.
Therefore, we use the following indicators to quantify the precision of our results:
\begin{align}
\label{eq_chi}
\chi^2_{AA}&=\frac{1}{4N}\sum_i(\Gamma_{\rm lt.}^\perp(\mu_0)^2+\Gamma_{\rm lt.}^\perp(p_i)^2)\left(\frac{1}{\Gamma_{\rm 
lt.}^\perp(p_i)}-\frac 1{\Gamma_{\rm th.}^\perp(p_i)}\right)^2\nonumber\\
\chi^2_{c\cb}&=\frac{1}{4N}\sum_i( J^{-2}_{\rm lt.}(\mu_0)+ J^{-2}_{\rm lt.}(p_i))
\left(J_{\rm lt.}(p_i)-J_{\rm th.}(p_i)\right)^2\nonumber\\
\end{align}
It corresponds to a sort of average between the (normalized) absolute error and the relative error.
In order to chose the parameters we took a value that gives a compromise between optimal values for both propagators.
We analyzed the errors for various values of $\alpha$ in both schemes both in $d=4$ and in $d=3$. For moderate values
of $\alpha$ (between 1 and 3) we do not observe an important dependence on $\alpha$. In the VM scheme, if the parameter
$\alpha$ is too small, the system shows a IR Landau pole and the results do not fit well the lattice data. For that
scheme we chose $\alpha=1$ and all results presented in that scheme corresponds to that value. 
 In the IS scheme, there is no Landau pole. For the $d=4$ case, the curves are almost insensitive to $\alpha$, even when it tends to 0.
 In the $d=3$ case, the best fits are obtained for $\alpha=0$. In the following, all our results in the IS  scheme are given for $\alpha=0$.
The corresponding values are presented in Tables~\ref{tab_best_fit} and \ref{tab_best_fit_3d}.

\begin{figure}[htbp]
 \includegraphics[width=4.2cm]{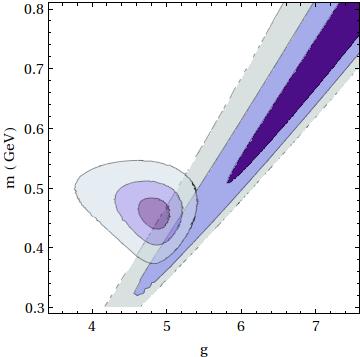}
 \includegraphics[width=4.2cm]{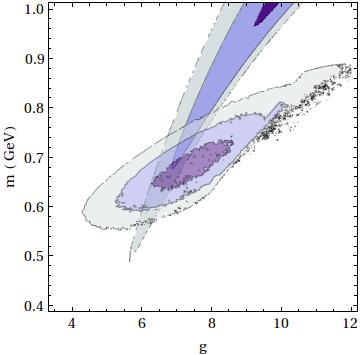}\\
 \includegraphics[width=4.2cm]{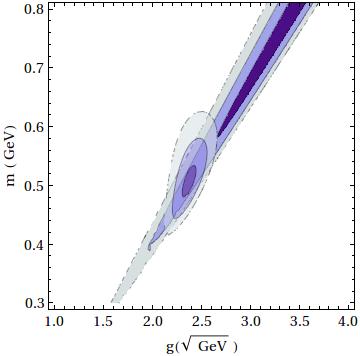}
 \includegraphics[width=4.2cm]{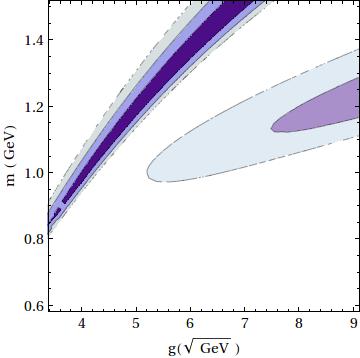}\\
 \caption{Contour levels for the quantities $\chi_{AA}$ and $\chi_{c\bar c}$ for the IS scheme (left) and VM scheme (right), both for $d=4$ (above) and $d=3$ (below).
 The large diagonal region corresponds to chi cc and the small elliptic one to chi AA.
 From dark to light: 4\%, 7\% and 10\%.}
\label{errorcontourlevels}
 \end{figure}

As our 1-loop expressions are certainly not exact
and given a certain tolerance of the estimate of the propagators, there are many possible values of the associated parameters.
In Fig.~\ref{errorcontourlevels} we present the contour levels associated to errors 4, 7 and 10\% for the quantities $\chi_{AA}$ and $\chi_{c\cb}$
both in $d=4$ and $d=3$ in schemes IS and VM. We can see that there is a region of acceptable parameters (with errors in both
2-point functions lower than 10\%) in almost all cases. The only exception is the VM scheme in $d=3$. In that dimension, the IS scheme
is much more precise than the other (and in fact gives an excellent fit for both 2-point functions simultaneously).
The same observation is also true for the 3-point functions discussed below. 
We observe that there is a large degeneracy of possible acceptable values for the parameters compatible with lattice data for
the ghost propagator. On the contrary, fitting the gluon propagator is much more demanding and the region of acceptable
parameters is much smaller.
\begin{table}[htbp]
 \begin{tabular}{|l|r|r|r|}
 \hline
  Scheme&$\alpha$&$g_0$&$m_0$ (GeV)\\
 \hline
 \hline
  IS&0.0& 5.2&0.44\\
 \hline
  IS&1.0& 5.2&0.43\\
 \hline
  IS&2.0& 5.8&0.48\\
 \hline
  IS&3.0& 6.3&0.53\\
 \hline
  VM&1.0& 7.5&0.77\\
 \hline
  VM&2.0& 9.0&0.78\\
 \hline
  VM&3.0& 9.1&0.75\\
 \hline
 \end{tabular}
\caption{Fitting parameters retained for computing correlation functions in $d=4$ for different schemes.}
\label{tab_best_fit}
\end{table}

\subsection{$d=4$}
We first present the results for the gluon propagator and the ghost dressing 
function. The $\beta$ functions were integrated with initial condition at 
$\mu_0=1$~GeV and we used the values of the mass and coupling constants given in Table I.

For all these schemes, we find a good agreement with the lattice data, with an 
error in-between 5 and 10$\%$ for $\chi_{AA}$ and $\chi_{\cb c}$. However, the 
ghost-antighost-gluon vertex functions are best reproduced with the VM scheme. 
We also present curves with the IS scheme. 
The difference between the two sets of curves gives an indication of the error 
of our calculation. 

\begin{figure}[htbp]
 \includegraphics[width=\linewidth]{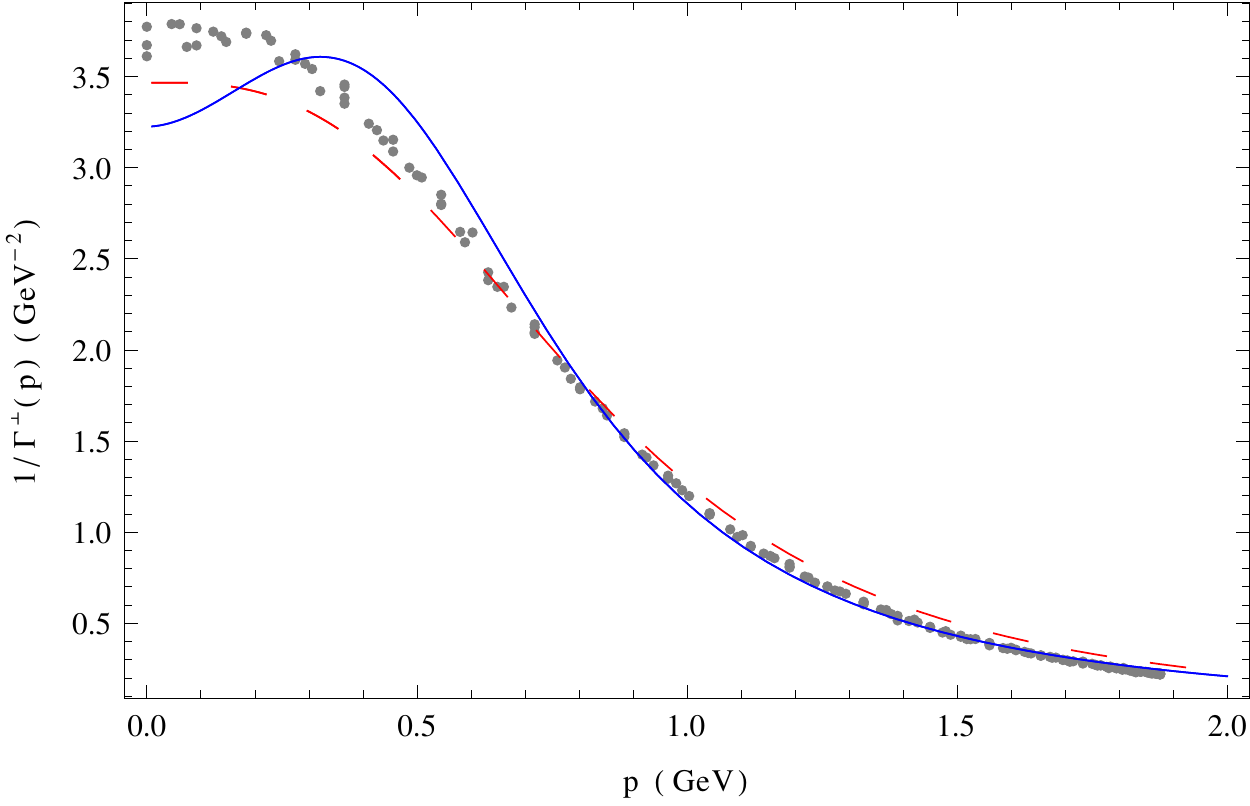}
 \includegraphics[width=\linewidth]{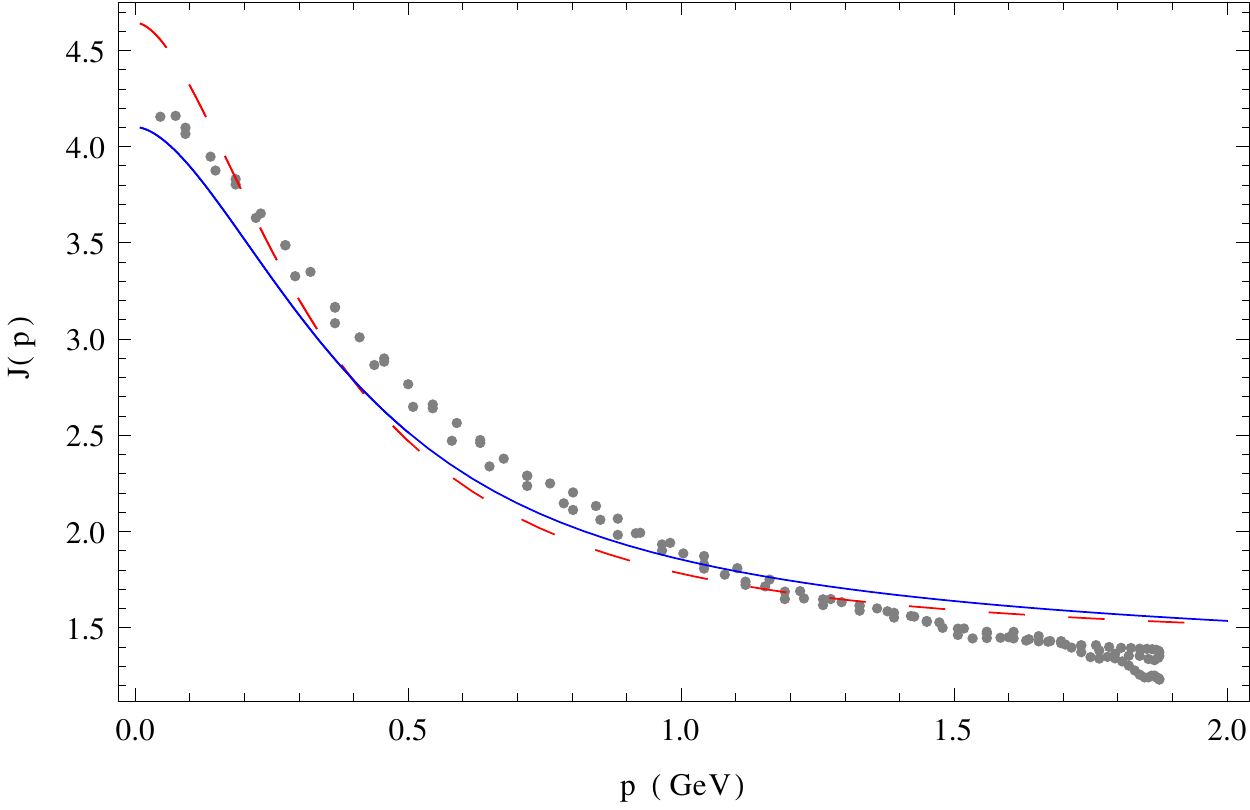}
\caption{Gluon propagator (top) and ghost dressing function 
(bottom) as a function of momentum in $d=4$. The points are lattice 
data of \cite{Cucchieri_08b}. The plain line (color online 
blue) corresponds to the infrared safe scheme with $\alpha=0$;
the dashed line (color online red) corresponds to 
the vanishing momentum scheme with $\alpha=1$.}
\label{fig_propags_4d}
 \end{figure}

\begin{figure}[htbp]
 \includegraphics[width=\linewidth]{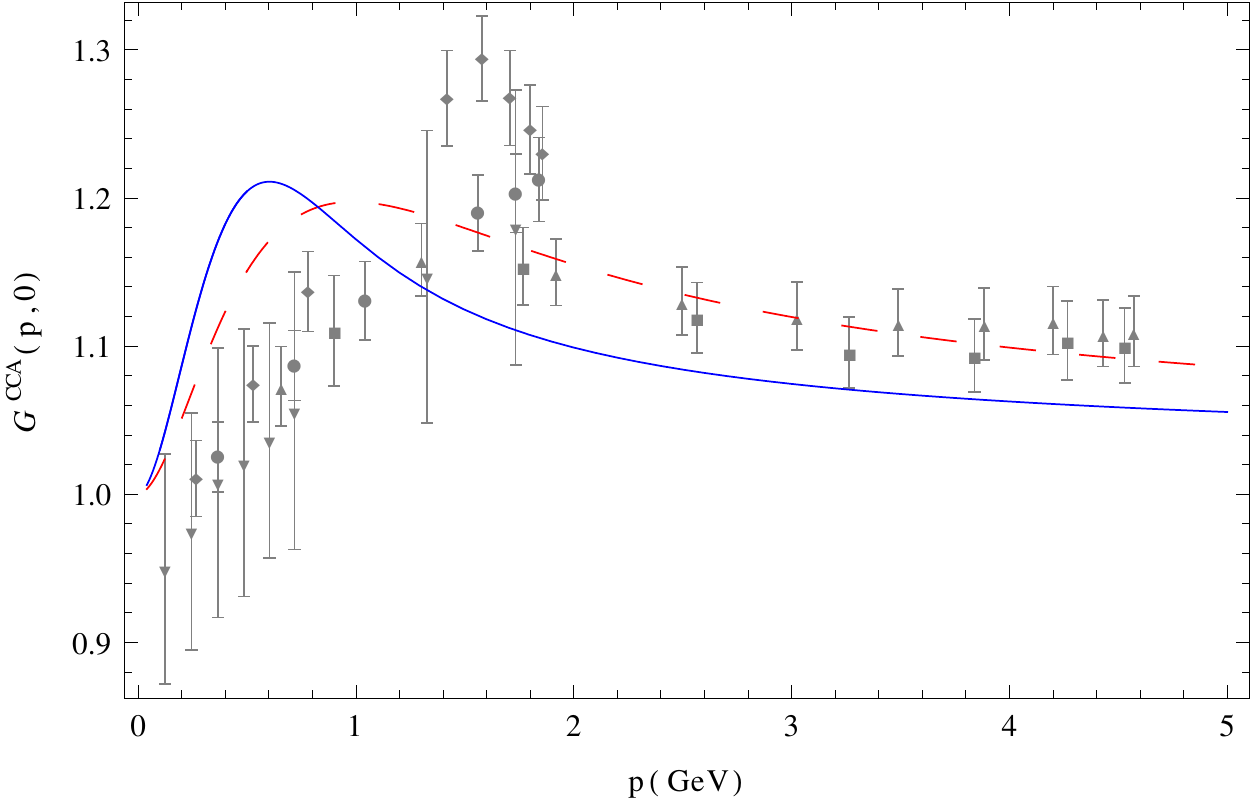}
 \includegraphics[width=\linewidth]{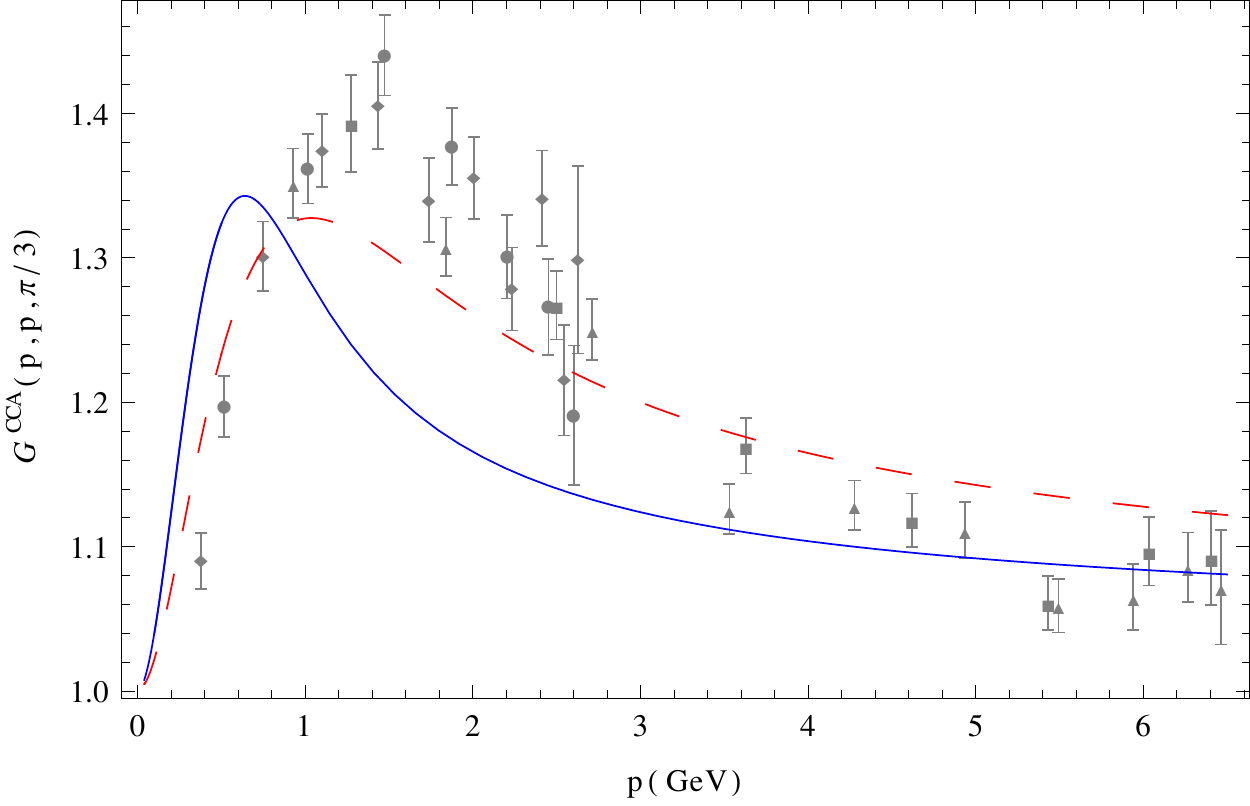}
 \includegraphics[width=\linewidth]{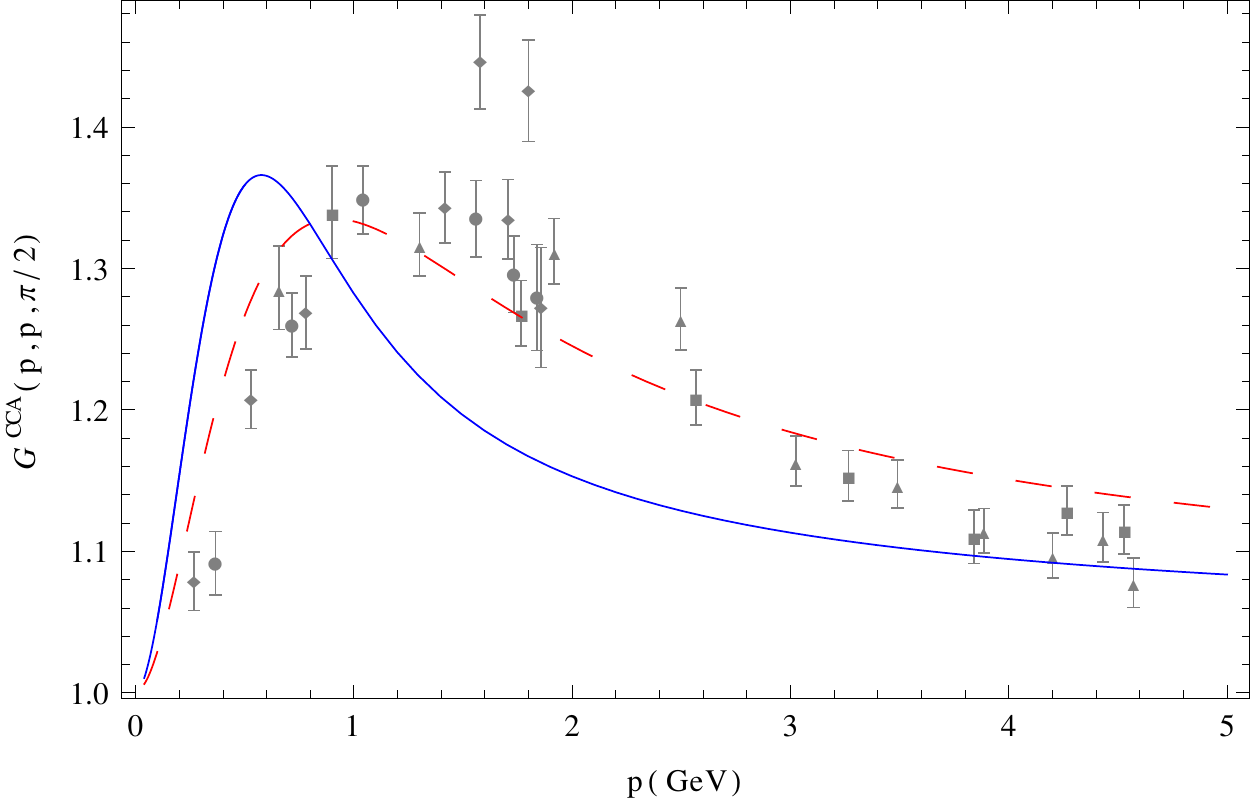}
\caption{Ghost-antighost-gluon correlation function $G^{c\cb A}$ for one vanishing momentum (top figure),
all momenta equal (middle figure), two momenta orthogonal, of equal norm (bottom) as a function of momentum, in $d=4$.
The lattice data of [\onlinecite{Cucchieri08}] are compared with our calculations.
See caption of Fig.~\ref{fig_propags_4d} for the 
legend.}
\label{fig_cbcA_4d}
 \end{figure}

\begin{figure}[htbp]
 \includegraphics[width=\linewidth]{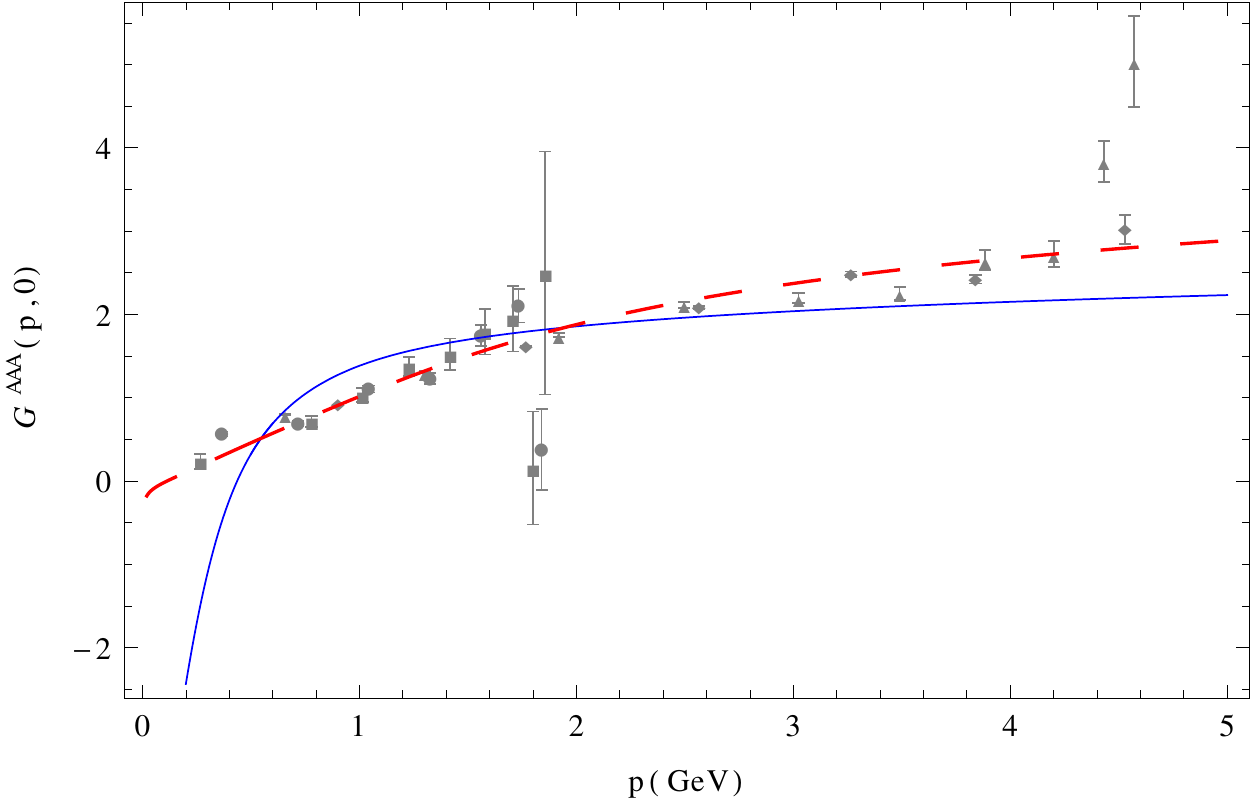}
 \includegraphics[width=\linewidth]{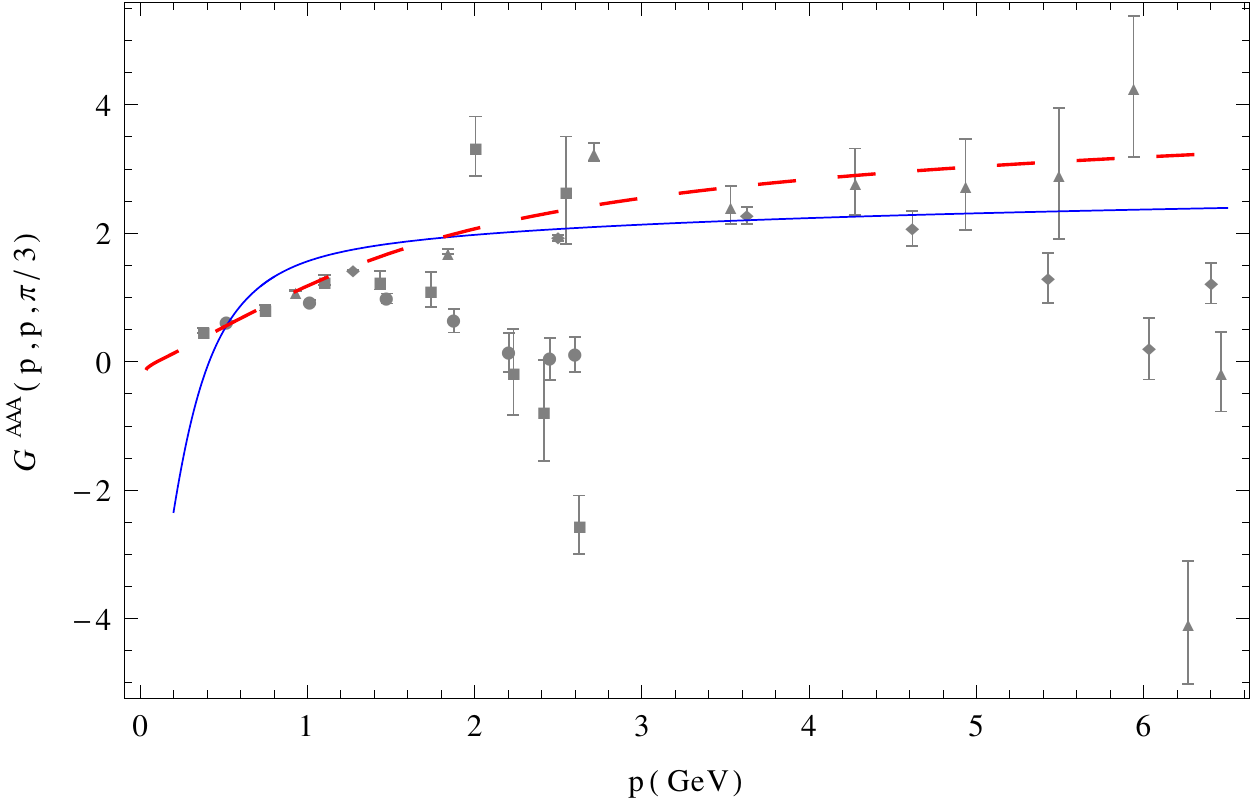}
 \includegraphics[width=\linewidth]{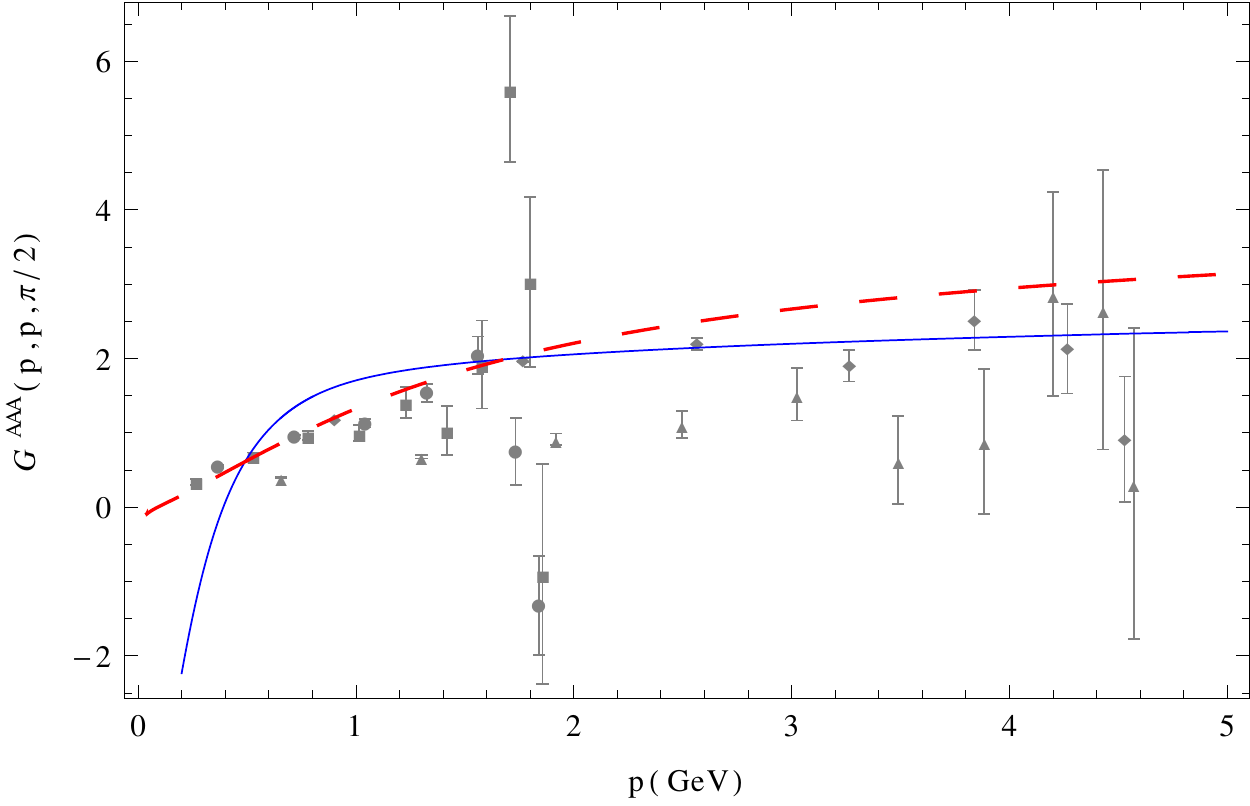}
\caption{Three gluon correlation function $G^{AAA}$ for one vanishing momentum (top figure),
all momenta equal (middle figure), two momenta orthogonal, of equal norm (bottom) as a function of momentum, in $d=4$. 
The 
lattice data of [\onlinecite{Cucchieri08}] are compared with our calculations. 
See caption of Fig.~\ref{fig_propags_4d} for the 
legend.}
\label{fig_AAA_4d}
 \end{figure}

The gluon propagator and ghost dressing functions are depicted in 
Fig.~\ref{fig_propags_4d}. 
We show in Fig.~\ref{fig_cbcA_4d} our results for $G^{c\cb A}$ and in
 Fig.~\ref{fig_AAA_4d} for $G^{AAA}$ for different momentum
 configurations.  

 In all cases, the results are compared with the
 corresponding results obtained in lattice simulations
 \cite{Cucchieri08}. The agreement is excellent.
 It is a
   striking result that the set of parameters adapted for describing
   the 2-point correlation function gives simultaneously a good
   agreement for the 3-point functions. When comparing the
   results with lattice data, it is important to note that the data
 for the 3-point functions (particularly for the $G^{AAA}$ functions)
 have large statistical errors and a full analysis of systematic
 errors has not been done yet. In consequence we can not completely neglect
 the errors coming from the lattice data with respect to those coming
 from the present calculation.

 \subsection{$d=3$}
\begin{figure}[htbp]
 \includegraphics[width=\linewidth]{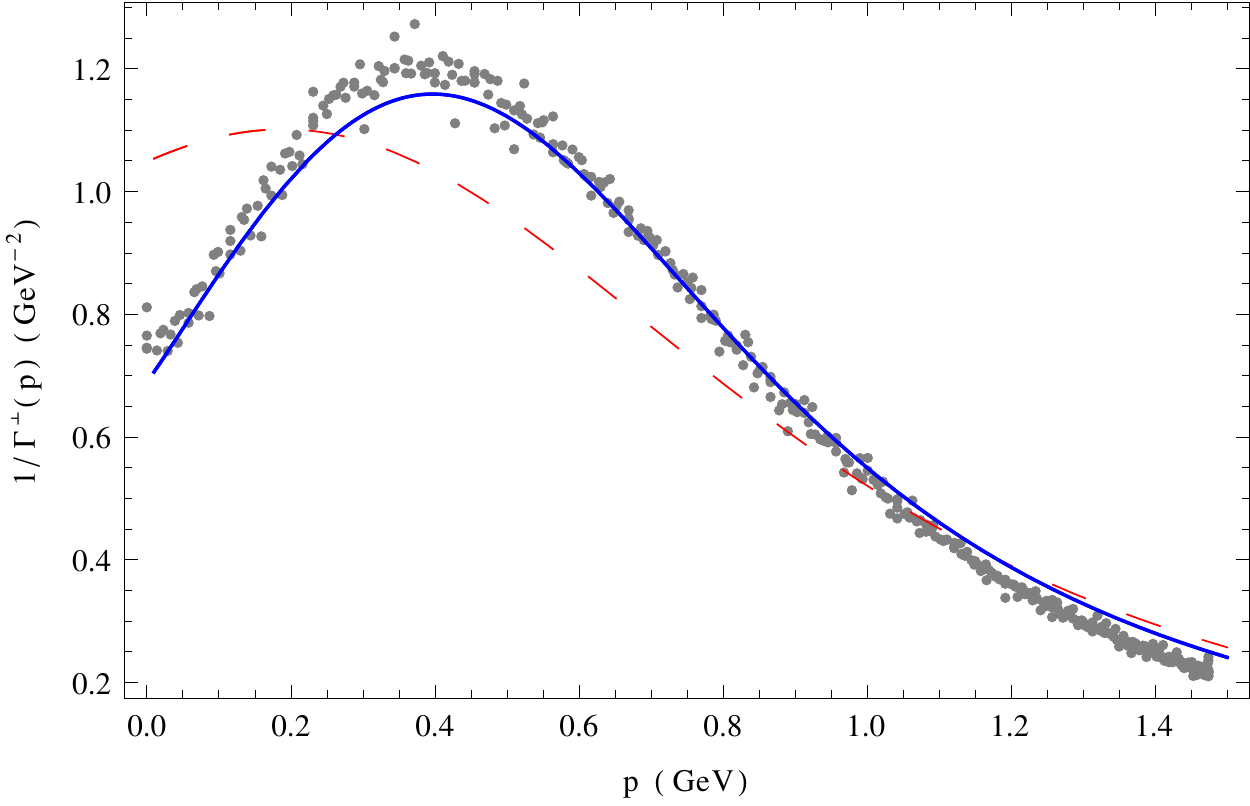}
 \includegraphics[width=\linewidth]{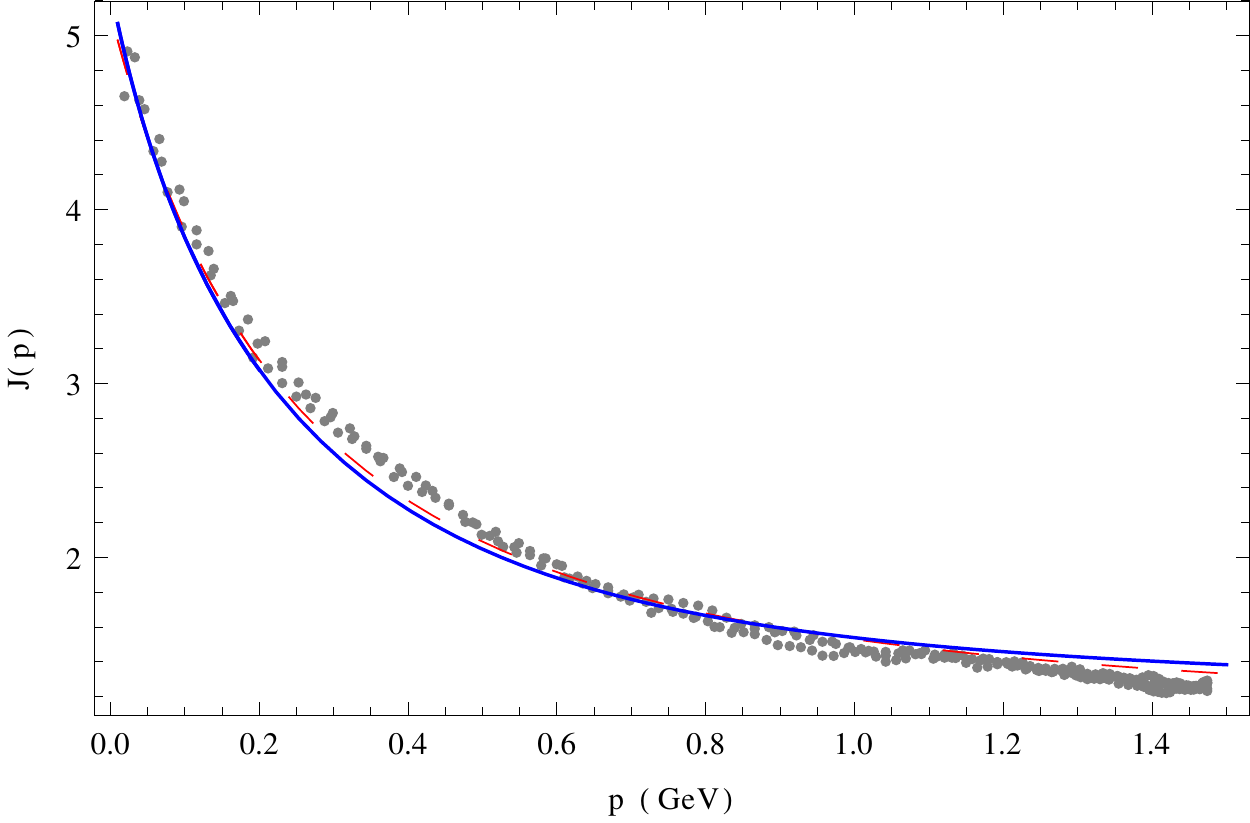}
\caption{Gluon propagator (top) and ghost dressing function 
(bottom) as a function of momentum in $d=3$. The points are lattice 
data of \cite{Cucchieri_08b}. The plain line (color online 
blue) corresponds to the infrared safe scheme with $\alpha=0$;
the dashed line (color online red) corresponds to 
the vanishing momentum scheme with $\alpha=1$.}
\label{fig_propags_3d}
 \end{figure}

A similar analysis can be performed in $d=3$.
We summarize in Table~\ref{tab_best_fit_3d} the parameters retained for the different renormalization schemes.
\begin{table}[htbp]
 \begin{tabular}{|l|r|r|r|}
 \hline
  Scheme&$\alpha$&$g_0$ (GeV$^{1/2}$)&$m_0$ (GeV)\\
 \hline
 \hline
  IS&0.0& 2.4&0.55\\
 \hline
  IS&1.0& 2.5&0.55\\
\hline
  IS&2.0& 2.5&0.55\\
 \hline
  IS&3.0& 3.0&0.65\\
 \hline
  VM&1.0& 4.0&1.00\\
 \hline
  VM&2.0& 4.5&0.95\\
 \hline
  VM&3.0& 6.1&1.11\\
 \hline
 \end{tabular}
\caption{Fitting parameters retained for computing correlation functions in $d=3$ for different schemes.}
\label{tab_best_fit_3d}
\end{table}
As explained before, the VM scheme does not give
results with good simultaneous agreement with lattice data for both propagators.
However, there is an excellent agreement in the IS scheme (with
errors in-between 5 and 10$\%$ for $\chi_{AA}$ and $\chi_{\cb c}$). In particular
the IS scheme correctly reproduces the increase of the gluon propagator
at low momentum. The best choice for $\alpha$ is zero.
We do not have a solid argument explaining why the
preferred renormalization scheme is different in $d=4$ and in
$d=3$. The gluon propagator and ghost dressing functions are depicted
in Fig.~\ref{fig_propags_3d}. The agreement remains very
good and, moreover, all qualitative aspects of the curves are
correctly reproduced. In particular, we find
  that the $G^{AAA}$ becomes negative at small momenta
  and diverges for vanishing momenta, in agreement with lattice
  results \footnote{$G^{AAA}$ also becomes negative in $d=4$ but at scales much smaller, probably difficult to observe in the lattice.
  More precisely, we obtain a divergence as $1/p$ in $d=3$ but only as  $\log(p)$ in $d=4$.}.

\begin{figure}[htbp]
 \includegraphics[width=\linewidth]{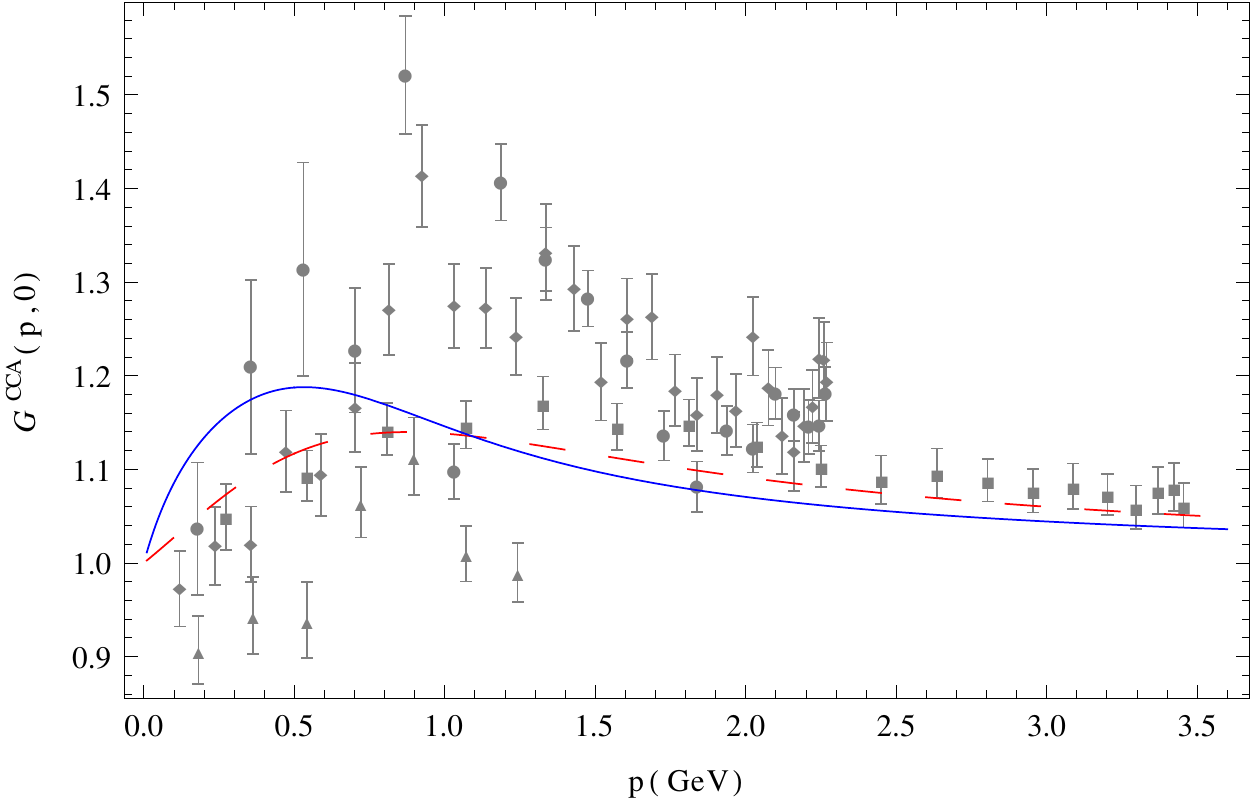}
 \includegraphics[width=\linewidth]{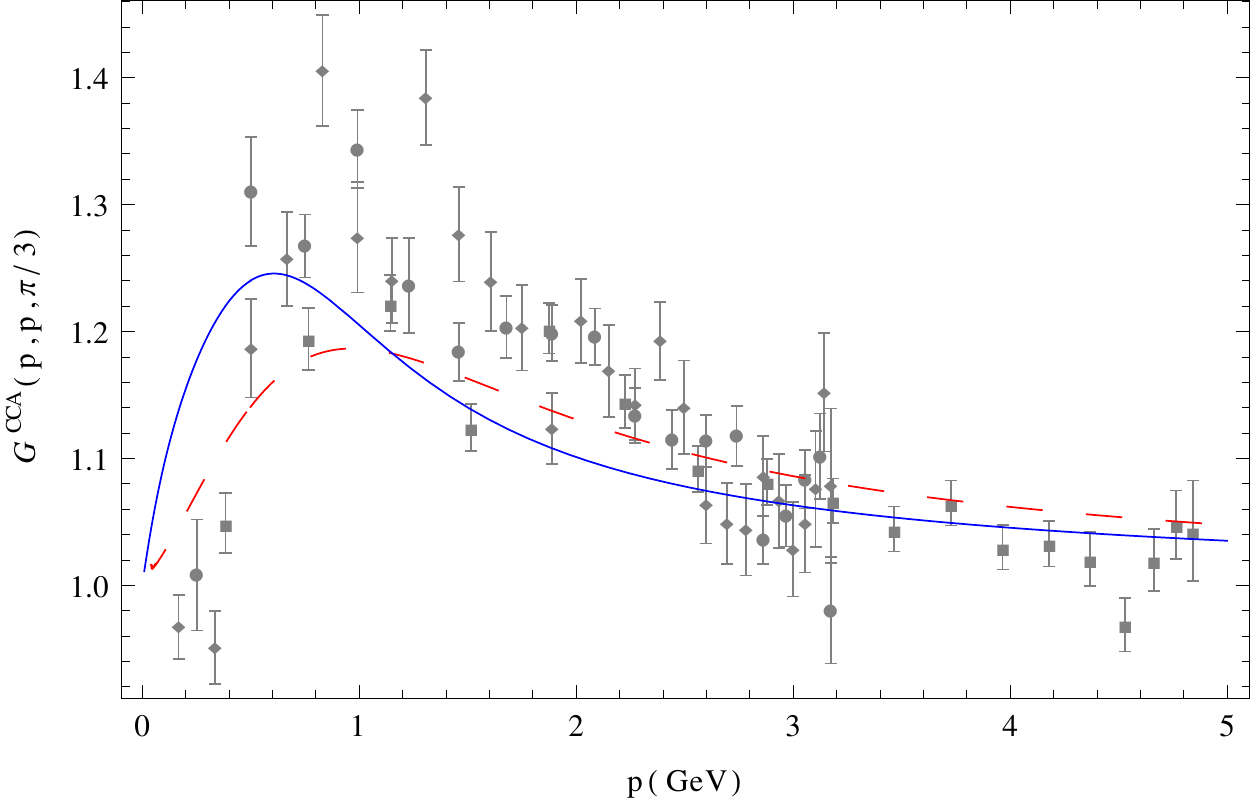}
 \includegraphics[width=\linewidth]{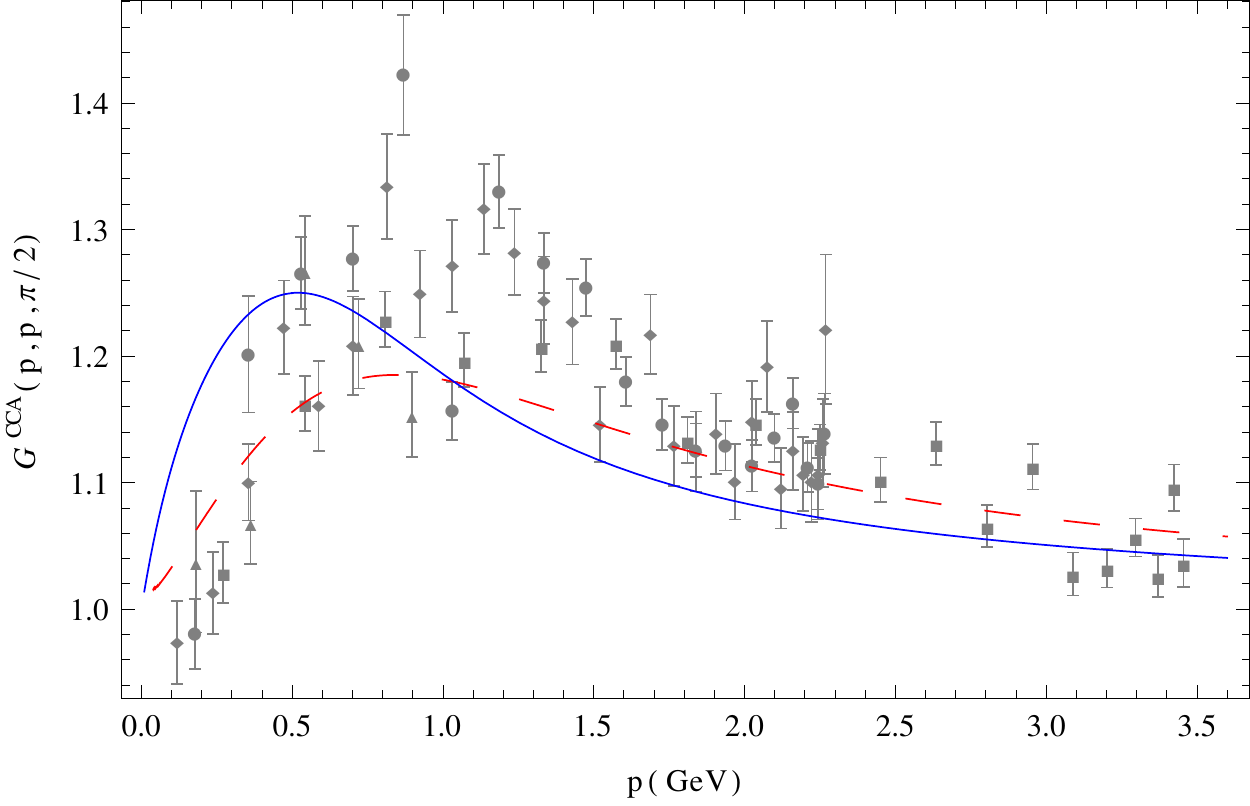}
\caption{Ghost-antighost-gluon correlation function $G^{c\cb A}$ for one vanishing momentum (top figure),
all momenta equal (middle figure), two momenta orthogonal, of equal norm (bottom) as a function of momentum, in $d=3$.
The lattice data of [\onlinecite{Cucchieri08}] are compared with our calculations.
See caption of Fig.~\ref{fig_propags_3d} for the 
legend.}
\label{fig_cbcA_3d}
 \end{figure}

\begin{figure}[htbp]
 \includegraphics[width=\linewidth]{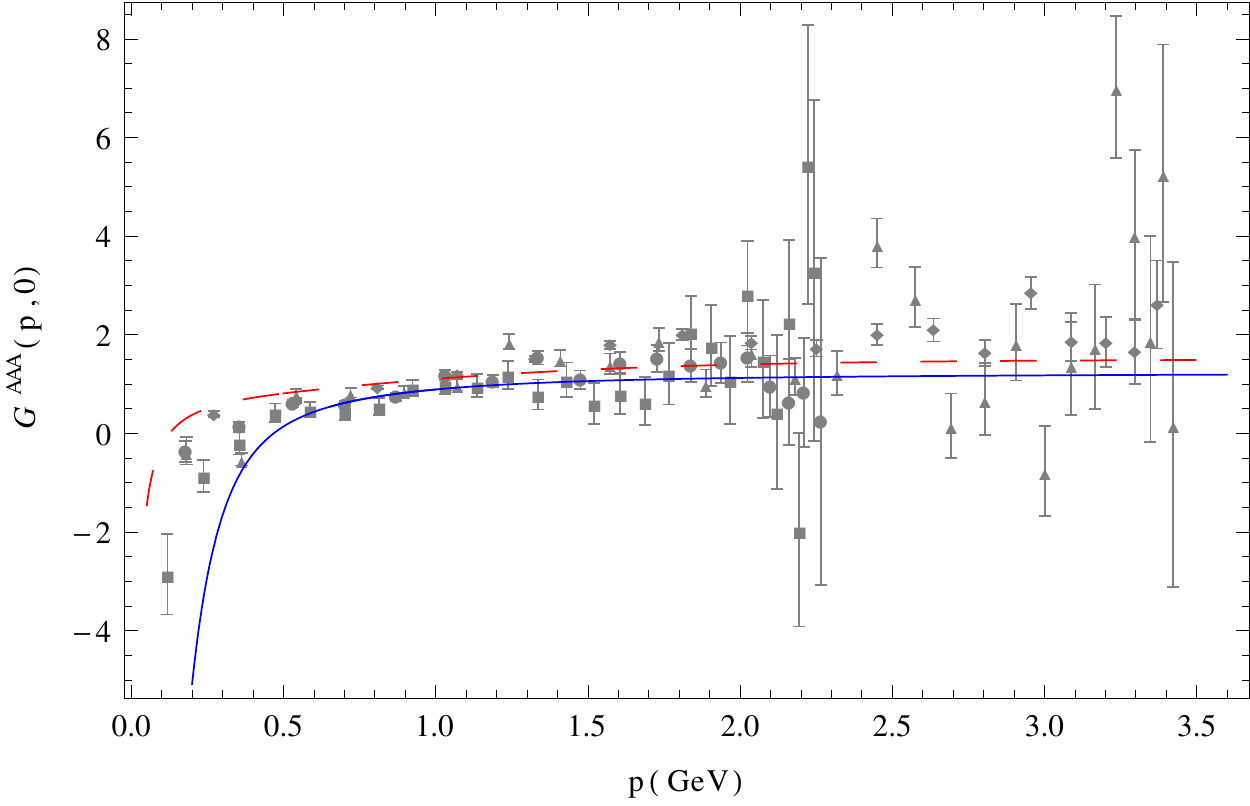}
 \includegraphics[width=\linewidth]{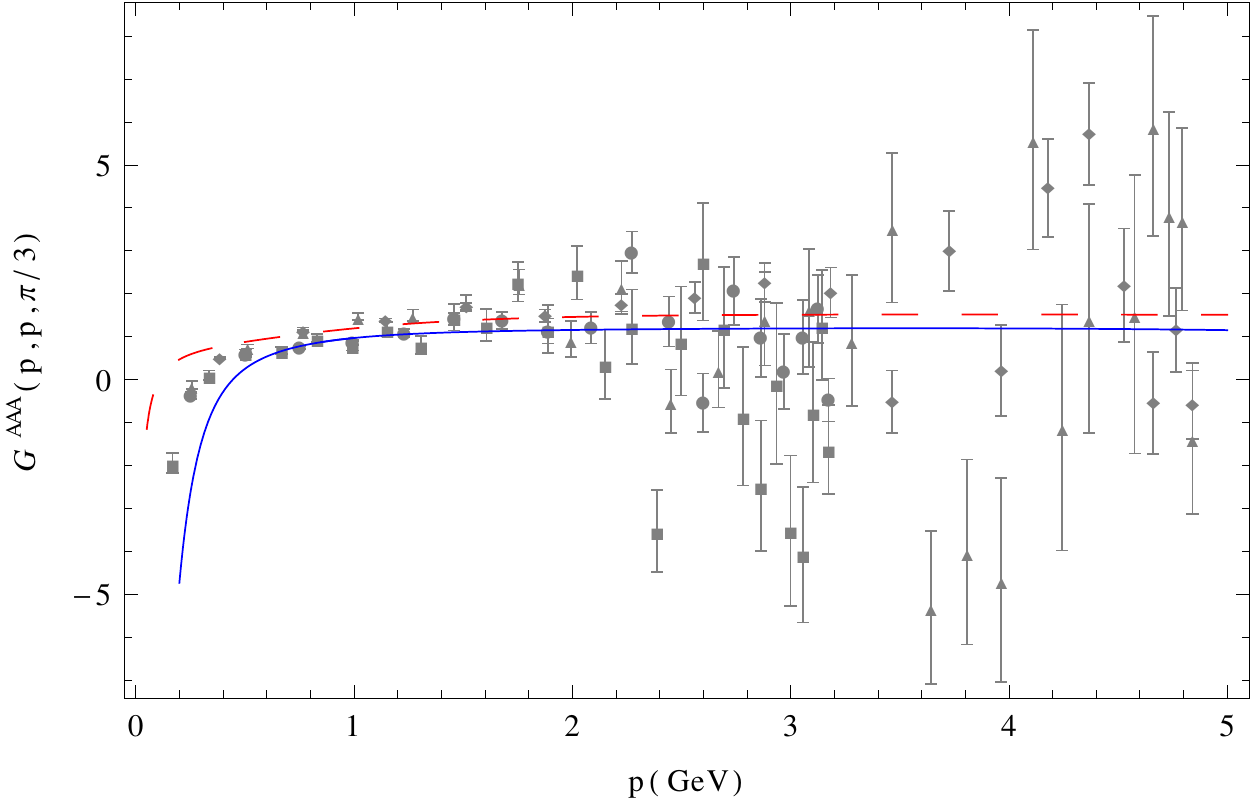}
 \includegraphics[width=\linewidth]{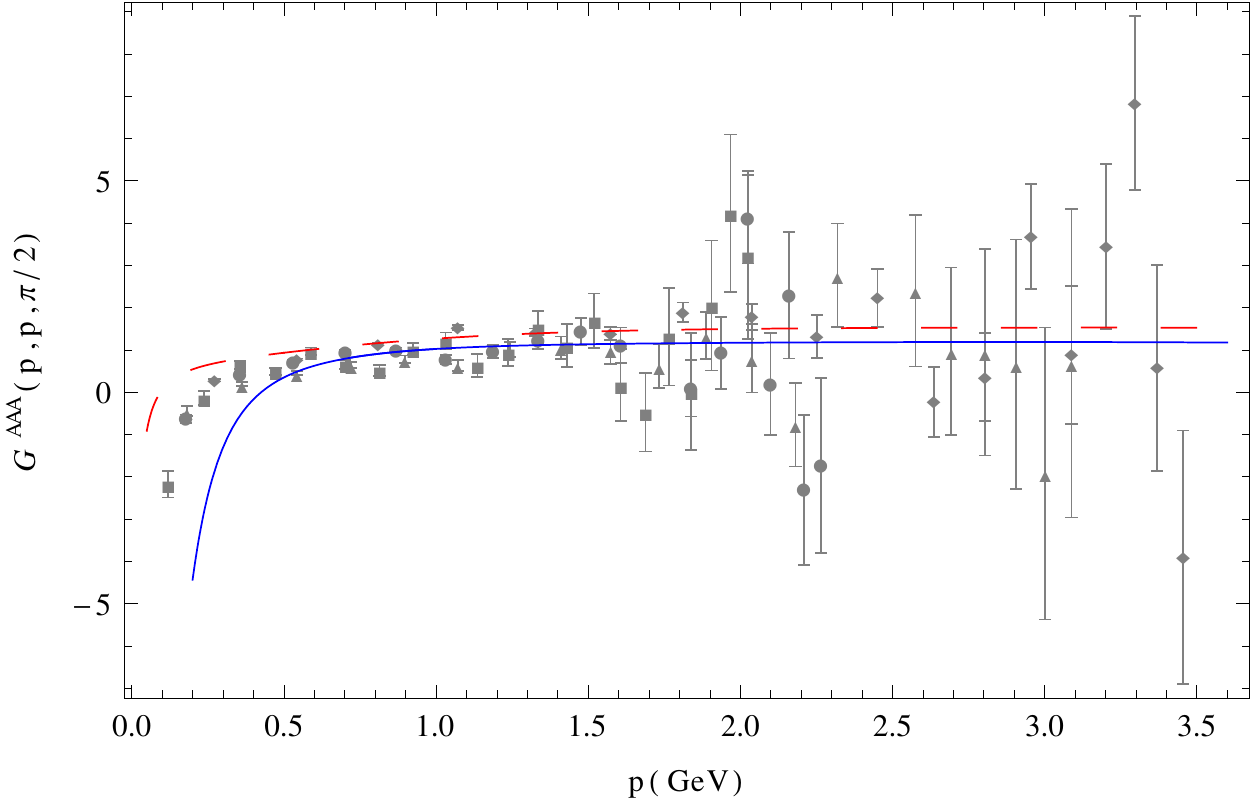}
\caption{Three gluon correlation function $G^{AAA}$ for one vanishing momentum (top figure),
all momenta equal (middle figure), two momenta orthogonal, of equal norm (bottom) as a function of momentum, in $d=3$. The 
lattice data of [\onlinecite{Cucchieri08}] are compared with our calculations.
See caption of Fig.~\ref{fig_propags_3d} for the legend.}
\label{fig_AAA_3d}
 \end{figure}

 We show in Fig.~\ref{fig_cbcA_3d} our results for $G^{c\cb A}$ and in
 Fig.~\ref{fig_AAA_3d} for $G^{AAA}$ for different momentum
 configurations. As for the four dimensional case, we used the parameters
 that lead to the best fits for the 2-point functions as inputs in our calculation of the 3-point functions.
 Consequently
 those functions are calculated without any free
 parameter (with the exception of the renormalization factor for the $G^{AAA}$ function mentioned previously).
 We obtain a very good agreement as in $d=4$.

\section{Conclusions}
\label{sec_conc}
In the present article we have presented a perturbative calculation of
3-point correlation functions in Landau-gauge, Yang-Mills theories in
$d=4$ and $d=3$ in all momentum regimes including the infrared.  Very
few analytical results were known up to now for infrared behavior of
the ghost-antighost-gluon vertex
\cite{RodriguezQuintero:2011au,Dudal:2012zx,Huber:2012kd,Aguilar:2013xqa}. For
the 3-gluon functions only educated ansatzes have been proposed
previously (see \cite{Huber:2012kd} and references therein).
Following \cite{Tissier:2010ts,Tissier:2011ey}, we introduced a bare
gluon mass so as to obtain controlled perturbative expressions both in
the ultraviolet and in the infrared regime and gives an infrared-safe
perturbative expansion for non-exceptional momentum configurations for
all $d>2$. Note that, in the gauge-fixing procedure of
\cite{Serreau:2012cg}, the mass-term naturally appears in the process
of lifting of the Gribov ambiguity.

We fixed the parameters of
the model by fitting the 2-point function to lattice simulations and by using two families
of renormalization group schemes. The resulting parameters are
  then used to calculate the 3-point functions. The comparison of the
  resulting functions with lattice simulations is therefore performed
 without any extra free-parameter (with the only exception of a renormalization factor in the $G^{AAA}$ function)
 and is very good. First, all qualitative
properties observed in lattice correlators are correctly
explained in a simple way. For example, it is observed in
lattice simulations that the 3-gluon correlator becomes negative in
$d=3$ at low momenta and even seems to diverge when all momenta go to
zero. This is a consequence of the IR divergence of the diagrams with ghost loops when all momenta go to zero. 
Second, not only the qualitative
agreement is very good, but also the comparison with lattice simulations gives an excellent quantitative agreement.
Note that the available lattice data still have large statistical errors and more precise results
(possibly with other tensor structures) would be welcome to give sharper test of our findings.

All these results strongly support the idea that at least an important
part of the infrared effects present in Quantum Chromodynamics can
take its origin in the Gribov-copies as has been suggested in last
years \cite{Gribov77,Zwanziger89,Zwanziger92,Dudal08}.  Contrarily to
previous analysis of these effects, the present approach does not
require the introduction of extra fields and the Feynman rules remain
almost identical to those of the standard perturbative analysis. Only
the gluon mass parameter is new and the calculation of many
correlators become treatable in practice, as shown in the present
article. Consequently, these studies can be extended in many aspects.
We are currently considering also the introduction of quarks that can
be done very easily. We also started studying the influence of higher
loop contributions to the 2-point correlation functions.

Of course, there are many open questions to be analyzed in the future. First of all, the inclusion of a mass term violates the nilpotency of standard BRST
transformations. Even if this difficulty is present in all approaches that take into account in various ways Gribov-copies effects, it has major consequences.
In particular, standard definition of the physical space of non-abelian gauge theories, based in the cohomology of the BRST charge is no longer applicable.
A new definition of the physical space is then required in order to be able to control the unitarity of the S matrix on it. This is a major problem that clearly goes
beyond the present article and that we would like to consider in the future.

\begin{acknowledgments}
The authors want to acknowledge A.~Maas for kindly making available the lattice data and for useful exchanges. They want also
acknowledge M.~Q.~Huber and J.~A.~Gracey for useful suggestions.
 The authors want acknowledge partial support from PEDECIBA and ECOS programs. N.~W. wants to thanks the LPTMC (UPMC), where
 most of the present work has been done, for its hospitality.
\end{acknowledgments}

\appendix

\section{BRST symetry}

In this appendix, we derive several constraints on the vertex
functions that can be deduced from the symmetries of the theory. We
recall that the action is invariant under the
Becchi-Rouet-Stora-Tiutin (BRST) transformation:
\begin{equation}
  \begin{split}
  \delta A_\mu^a&=\zeta(\partial_\mu c^a+g_0 f^{abc}A_\mu^b c^c),\\
\delta c^a&=\zeta(-\frac 12 g_0 c^bc^c),\\
\delta\cb^a&=\zeta ih^a,\qquad
\delta ih^a=\zeta m_0 c^a.    
\end{split}
\end{equation}
where $\zeta$ is a Grassmann parameter.  As usual, we introduce sources for the BRST variations of the fields $A$ and $c$ with the 
following action:
\begin{equation}
  \label{eq_Ssources}
  \begin{split}
  S_{\rm sources}=\int_x \Big[\bar K_\mu^a&\left(\partial_\mu 
c^a+g_0f^{abc}A_\mu^b c^c\right)\\&-\frac {g_0}2f^{abc}\bar L^ac^bc^c\Big].
  \end{split}
\end{equation}
The Slavnov-Taylor (ST) identity associated with this symmetry reads:
\begin{equation}
\label{eq_BRST_ST}
  \begin{split}
\int d^dx \bigg\lbrace \frac{\delta \Gamma}{\delta A_\mu^a}\frac{\delta 
\Gamma}{\delta \bar K_\mu^a} &+ 
\frac{\delta \Gamma}{\delta c^a}\frac{\delta \Gamma}{\delta \bar L^a} - 
ih^a\frac{\delta\Gamma}{\delta\bar c^a} \\&+ 
im_0^2 \frac{\delta\Gamma}{\delta h^a}c^a\bigg\rbrace = 0    
  \end{split}
\end{equation}
We will also make use of the symmetry: 
\begin{equation}
  \label{eq_t}
  \begin{split}
\delta A_\mu^a&=\delta c^a=0,\qquad  \delta \cb^a=\epsilon c^a, \\ 
\delta ih^a&=-\epsilon \frac{g_0}{2}f^{abc}c^bc^c.
  \end{split}
\end{equation}
where $\epsilon$ is an infinitesimal real number. 
 The associated ST identity reads:
\begin{equation}
\label{eq_t_ST}
\int d^dx \left\lbrace c^a\frac{\delta\Gamma}{\delta \bar 
c^a}+i\frac{\delta\Gamma}{\delta h^a}\frac{\delta\Gamma}{\delta \bar 
L^a}\right\rbrace=0
\end{equation}
Finally, we can write a Ward identity associated with the invariance
of the theory under an infinitesimal shift in the antighost $\cb(x)\to \cb(x)+\epsilon(x)$,
which reads:
\begin{equation}
  \partial_\mu\frac{\delta\Gamma}{\delta\bar 
K_\mu^a(x)}=\frac{\delta\Gamma}{\delta\bar c^a(x)}
\end{equation}
Deriving this equation once with respect to $c$ and taking the Fourier 
transform, we find:
\begin{equation}
\label{eq_kbccbc}
-i p_\mu \Gamma^{(2)}_{c^b\bar K_\mu^a}(p)=\Gamma^{(2)}_{c^b\bar c^a}(p)  
\end{equation}
Deriving once more with respect to $A$ and taking the Fourier
transform, we obtain:
\begin{equation}
\label{eq_kbcAcbcA}
 -i p_\mu \Gamma^{(3)}_{c^a\bar K_\mu^b A_\nu^c}(p,k,r)=\Gamma^{(3)}_{c^a\bar 
c^b A_\nu^c}(p,k,r) 
\end{equation}
This last expression justifies the tensorial decomposition
(\ref{eq_vertex_cbcA_def}) and shows that:
\begin{equation}
  \Gamma^{(3)}_{c^a\bar K^b_\nu A^c_\mu}(p,k,r)=-igf^{abc}\Gamma_{\nu\mu}(p,k,r)
\end{equation}
if we suppose a color structure proportional to $f^{abc}$, as was done all along this article.

We can now prove Eq.~(\ref{eq_BRST_constraint}) by deriving
Eq.~(\ref{eq_BRST_ST}) with respect to $\cb$ and twice with respect to
$A$ and expressing the vertex involving $\bar K$ by using
Eqs.~(\ref{eq_kbccbc},\ref{eq_kbcAcbcA}).

Eq.~(\ref{eq_non_renorm}) is obtained much in the same way. we first
derive Eq.~(\ref{eq_BRST_ST}) with respect to two ghost and one
antighost fields, and Fourier transform. We thus get: 
\begin{equation}
  \begin{split}
  &-\Gamma^{(3)}_{c^c\bar c^b A_\mu^d}(k,r,p)\Gamma^{(2)}_{c^a\bar 
K_\mu^d}(p)+\Gamma^{(3)}_{c^a\bar c^b A_\mu^d}(p,r;k)\Gamma^{(2)}_{c^c\bar K_\mu^d}(k)
\\&+\Gamma^{(2)}_{c^d\bar c^b}(r)\Gamma^{(3)}_{c^ac^c\bar L^d}(p,k;r)=0    
  \end{split}
\end{equation}
The vertex that involves $\bar L$ can be re-expressed by deriving
Eq.~(\ref{eq_t_ST}) with respect to $A$ and to $c$ twice, and Fourier
transforming:
\begin{equation}
  \Gamma^{(3)}_{c^d\bar c^c A_\mu^b}(p,k;r)-\Gamma^{(3)}_{c^c\bar c^d 
A_\mu^b}(k,p;r)+ir_\mu\Gamma^{(3)}_{c^dc^c\bar L^b}(p,k;r)=0
\end{equation}
The last two equations can be used to prove (\ref{eq_non_renorm}) using again, 
a color structure proportional to $f^{abc}$.


\begin{thebibliography}{10}
\bibitem{supplemental} 
Mathematica file with all our results for the
  scalar functions appearing in the two and three-point correlation functions.
\end{thebibliography}
\end{document}